\documentclass[aps,prd,showpacs,nofootinbib,floats,floatfix,preprintnumbers,groupedaddress,
]{revtex4}
\usepackage{comment}
\usepackage{graphicx}
\usepackage{amsmath, mathrsfs}
\usepackage{float}
\usepackage{subfigure}
\usepackage[usenames]{color}
\usepackage{amssymb}
\usepackage{bbm}
\usepackage{csquotes}
\usepackage{tensor}
\usepackage{hyperref}
\usepackage{url}
\usepackage{orcidlink}
\usepackage{tikz}
\usetikzlibrary{positioning,decorations.pathmorphing}

\usepackage{epstopdf}
\newcommand{\bea}{\begin{aligned}}
\newcommand{\eea}{\end{aligned}}
\newcommand{\beq}{\begin{equation}}
\newcommand{\eeq}{\end{equation}}

\newcommand{\bse}{\begin{subequations}}
\newcommand{\ese}{\end{subequations}}

\newcommand{\lie}{\pounds}

\definecolor{burgundy}{RGB}{139,26,26}
\usepackage{hyperref}
\hypersetup{
     colorlinks   = true,
     citecolor    = blue,
     linkcolor    = burgundy,
     urlcolor     = burgundy,
}

\newcommand{\bmm}{\begin{multline}}
\newcommand{\emm}{\end{multline}}

\begin{document}
\title{Lorentz charges for Lovelock gravity}
\author{Ayan Chatterjee\,\orcidlink{0000-0000-0000-0000}}
\email{E-mail: ayan.theory@gmail.com}
\affiliation{Department of Physics \& Astronomical Science, Central University of Himachal Pradesh, Dharamshala-
176215, India.}
\author{Sahil Devdutt\,\orcidlink{0009-0002-5887-4520}}
\email{E-mail: devduttsahil@gmail.com}
\affiliation{Department of Physics \& Astronomical Science, Central University of Himachal Pradesh, Dharamshala-
176215, India.}
\author{Amit Ghosh\,\orcidlink{0000-0000-0000-0000}}
\email{E-mail: amit.ghosh@saha.ac.in}
\affiliation{Theory Division, Saha Institute of Nuclear Physics, 1/AF Bidhannagar, Kolkata- 700064\\
and Homi Bhabha National Institute, Anushakti Nagar, Mumbai 400094, India.}
\author{Avirup Ghosh\,\orcidlink{0000-0000-0000-0000}}
\email{E-mail: avirup.avi@gmail.com}
\affiliation{Kulti College, Paschim Bardhaman, West Bengal- 713343, India.}
\pagenumbering{arabic}
\renewcommand{\thesection}{\arabic{section}}
\begin{abstract}
In this paper, we construct the charges associated with the residual local Lorentz transformations on a black hole horizon in the Lovelock class of theories of gravity. The horizons is assumed to be non- expanding horizons, \emph{i.e} a null hypersurface for which the expansion corresponding to the null generator vanish. Naturally, these hypersurfaces preserve only a subgroup of the full Lorentz group of the bulk, and the residual symmetries correspond to the little group which stabilizes the horizon null generator. We show that on the horizon, the Hamiltonian charge corresponding to the residual local Lorentz boost is the Wald entropy.   
\end{abstract}
\maketitle
\section{Introduction}
The laws of black hole mechanics as proposed in \cite{Bardeen:1973gs} point to the fact that horizons in asymptotically flat spacetimes are thermodynamic systems. Subsequently, the thermal nature of horizons have been extended to spacetimes with different asymptotics as well. For example, black holes in de-Sitter spacetimes have a dual thermal character arising from thermality of black holes as well as the cosmological horizons \cite{Gibbons:1977mu}. So the thermal nature of horizons is not restricted to black hole horizons, but rather is shared by cosmological horizons as well. Although the resemblance of laws of black hole mechanics and the laws of thermodynamics was initially viewed as formal analogy, but this is now considered  an exact correspondence after Hawking demonstrated that black holes can emit radiation \cite{Hawking:1974rv,Hawking:1975vcx}. Interestingly, the idea that black holes posses entropy was first proposed by Bekenstein. Guided by the second law of thermodynamics, he argued that black holes must have entropy related to the cross sectional area of the horizon \cite{Bekenstein:1972tm,Bekenstein:1973ur}. Subsequent semiclassical calculations of Hawking provided the requisite underpinning by showing that horizons may be assigned a temperature proportional to their surface gravity. Although surface gravity $\kappa$ and black hole temperature $T$ are not \textit{a priori} related, this calculation established their fundamental connection when one includes the quantum mechanical effects. In fact, classically one can argue that black holes have absolute zero temperature \cite{Bardeen:1973gs,Bekenstein:1973ur}. Similarly one should expect that the identification of black hole area with entropy may have reasons rooted in quantum mechanics. The earliest of such attempts to understand black hole entropy is due to Gibbons and Hawking, who developed the methods of Euclidean quantum gravity \cite{Gibbons:1976ue}. By evaluating the action for complexified black hole solutions which contributes to the logarithmic of the partition function they recovered the Bekenstein-Hawking formula for the entropy. Later on it was argued that the Euclidean methods of calculating the black hole entropy are equivalent to classical methods  \cite{Wald:1993nt}. A comparison between these two frameworks is given in \cite{Iyer:1995kg}. Another very significant approach is based on a proposal of relating black hole entropy with entanglement entropy associated with the quantum correlations among vacuum fluctuations between the inner and the outer regions of a black hole solution \cite{Bombelli:1986rw,Callan:1994py,Holzhey:1994we,Srednicki:1993im,Frolov:1993ym,tHooft:1984kcu}. This framework also successfully reproduces the fact that entropy scales with the area of the horizon with the dominant contribution to the horizon entropy arising due to short distance correlations of UV- regulated vacuum fluctuations across the horizon \cite{Solodukhin:2011gn}. Recently, the holographic approach has provided significant insights into the origin of entropy through the proposal of Ryu and Takayanagi, where they argued that the entanglement entropy of boundary theory can be related to minimal surfaces in the bulk \cite{Ryu:2006bv}. One may also obtain further discussions on these ideas in the context of time- dependent spacetimes as is discussed in \cite{Hubeny:2007xt}. For theories of gravity beyond the Einstein theory, generalisations of the holographic entropy proposal are given in \cite{Camps:2013zua, Dong:2013qoa, Faulkner:2013ana}. A review of these ideas are in \cite{Nishioka:2009un, VanRaamsdonk:2016exw}.\\

Despite the diversity of approaches, it is universally accepted that the definitive understanding of black hole entropy will come from the understanding of black hole horizon microstates.
These perspectives into the microscopic origins of entropy are provided within the framework of string theory and loop quantum gravity as well \cite{Strominger:1996sh,Rovelli:1996dv,Dijkgraaf:1996it,Ashtekar:1997yu,Kaul:1998xv,Peet:2000hn,Dabholkar:2004yr,Domagala:2004jt,Meissner:2004ju,Ghosh:2004wq,Dabholkar:2005dt,Engle:2009vc,Kaul:2000kf,Mandal:2010cj}. Such quantum states may only be reasonably and faithfully determined once a complete understanding of the quantum theory of gravity is obtained.
A quantum theory of gravity is yet to be constructed, and in the absence of such a theory, we must rely on effective quantum field theory and symmetry arguments to gain valuable insights into the fundamental physics of black hole or cosmological horizons. Although it remains unclear which of the symmetries of the classical theory of gravity may be implemented in the corresponding quantum theory of gravity, nevertheless, over the years, frameworks based on the symmetry principles have been successful and have led us towards the better understanding of black hole entropy in terms of its horizon area:
\cite{Brown:1986nw,Lee:1990nz,Wald:1993nt,Carlip:1994gy,Iyer:1994ys,Carlip:1993sa,Strominger:1997eq,Carlip:1999cy,Chatterjee:2020iuf,Garg:2026toh}. One important work based on symmetry arguments is due to Carlip \cite{Carlip:1999cy} in which he analysed the algebra of vector fields, and their corresponding Hamiltonian charges in the presence of Killing horizon and found that the Hamiltonian charge algebra acquires a central extension which contains the Virasoro sub-algebra. This is similar to the central extension of asymptotic symmetry algebra known in the context of black hole solutions in $3$ dimensions \cite{Brown:1986nw}. Using the Cardy formula \cite{Cardy:1986ie,Blote:1986qm} for asymptotic density of states, he showed that one obtains the Bekenstein-Hawking formula for the black hole entropy without the knowledge of the exact nature of underlying quantum theory of gravity. Similar result also hold true for black holes in $AdS_{3}$ spacetime \cite{Strominger:1997eq}.
However, such analysis are not without controversy, in particular, when the analysis was further carried out in the framework of isolated horizons and it arose that the central charge vanishes if one considers the symmetry structure intrinsic to the horizon. Although on the stretched horizon, 
the algebra survives, with the entropy differing from the usual results by a factor of $\sqrt{2}$ \cite{Dreyer:2001py}.  \\

Another attempt to understand horizon thermodynamics through symmetry arose through the Wald interpretation of black hole entropy as a Noether charge associated with the Killing symmetries of the horizon \cite{Wald:1993nt}. He showed that one may present a classical interpretation of black hole entropy as a consequence of residual symmetries generating the horizon. In particular, the black hole entropy arises as the Noether charge of the Killing vector field generating the Killing horizon. The ramifications of this classical definition of entropy was made clear through further studies where it was shown that the Wald entropy in the Lovelock class of theories of gravity does indeed behave like an entropic function, since this quantity always grows in a classical physical process \cite{Iyer:1994ys,Chatterjee:2011wj, Kolekar:2012tq,Sarkar:2013swa, Wall:2015raa}. It is worth noting that while Wald's original derivation assumes the presence of a bifurcate Killing horizon, Sen's entropy function formalism \cite{Sen:2005wa,Astefanesei:2006dd} provides a method to compute entropy of extremal black holes which do not have bifurcation surface. The question which remains is why should entropy be associated with symmetries at all? Is it a mere coincidence or should one reinterpret Wald entropy as the geometric area associated with the boost symmetries? Such a question is pertinent since in \cite{Banados:1993qp} Ba\~nados et al. had argued that in all dimensions the classical black hole area is canonical conjugate to the Euclidean deficit angle $\Theta$. The similar results were also obtained for all dimensions in \cite{Carlip:1993sa} and were discussed in details for $2+1$ dimensions in \cite{Carlip:1994gc}. Further, this deficit angle in Euclidean signature is related to boost in Lorentz signature \cite{Witten:2024upt}. Therefore geometric area of the black hole can be thought of conjugate to boost parameter $\eta$. A similar correspondence between the area of the bifurcation sphere and boost energy outside the Rindler horizon is dicussed in \cite{Wall:2010cj}. So the natural question would be to propose that such a consideration should be valid even for black holes in higher dimensions as well as in any diffeomorphism invariant theories of gravity. This question was raised in \cite{Chatterjee:2020iuf,Chatterjee:2015lwa} and answered in affirmative for isolated horizons in 4-dimensional GR. An extension of this framework to entire Lovelock class of theory in the context of the first order tetrad- connection formalism is the objective of this paper, although our approach is different from the one used in \cite{Banados:1993qp,Carlip:1993sa,Carlip:1994gc}.\\

Let us now precisely define the problem: In this work we are going to adopt symmetry based perspective and exploit the non-trivial behavior of gauge transformations in the presence of boundaries to establish that for any arbitrary diffeomorphism invariant theory of gravity belonging to the Lovelock class, the Hamiltonian charge corresponding to the boost symmetry \emph{on} the horizon is exactly the geometric function which is identical to the Wald entropy. So from this point of view Wald entropy is nothing but a modified geometric area of the horizon associated with boost symmetry. The idea stems from a very general observation that in gauge theories on a spacetime with boundaries, the boundary conditions play a crucial role in precisely identifying the allowed gauge transformations: Only those transformations in bulk survive on the boundary which preserve the boundary conditions. For example, in diffeomorphism invariant theories if one introduces a boundary $\partial \mathcal M$ on a manifold $\mathcal{M}$, then the set of vector fields generating infinitesimal diffeomorphisms at the boundary is restricted to contain only those vector fields which are tangent to $\partial\mathcal{M}$, since only those vector fields will preserve the location of the boundary. This in turn, renders certain gauge transformations physical which give rise to physical charges that may be used to label degrees of freedom on the boundary. The significance of such 
degrees of freedom in low dimensional physics has been 
known for quite sometime \cite{Zhang:1988wy,Frohlich:1990xz,Balachandran:1991dw,Wen:1992vi}. In gravity in 2+1 dimensions Carlip has shown that these degrees of freedom precisely account for black hole entropy \cite{Carlip:1994gy}. For more recent advancements see \cite{Freidel:2020xyx,Freidel:2020svx,Freidel:2020ayo}. In this work we are going to use the covariant phase space formalism which is well suited for the computation of conserved charges in relativistic field theories while preserving the diffeomorphism and Lorentz invariance \cite{Zuckerman:1986vzu,Witten:1986qs,Crnkovic:1987tz,Lee:1990nz,Wald:1993nt,ASHTEKAR1991417}. In this formalism the gauge transformations are generated by vector fields, called zero modes. These zero modes form a Lie-algebra and are generators of the continuous gauge group $G$ of the theory. Generally, these gauge directions are the degenerate directions of the symplectic form, and do not give rise to any physical charge. But in the presence of boundaries, these would be zero modes, become relevant and acquire non-vanishing physical charge. It is precisely these transformations, which would otherwise be zero modes, give rise to Wald entropy as physical charge on the horizon.\\

We shall argue in this paper that, for 
the Lovelock class of theories, the Wald entropy (modified geometric area) is the Hamiltonian charges corresponding to the residual local Lorentz boost symmetries on the horizon. More precisely, we shall show that, this connection between local Lorentz boost and Hamiltonian charges are not restricted only to Rindler like local observers who are off and far from the horizons but a character of the horizon itself and must be attributed as construction which is independent of the observer. Such a claim should not be surprising since it is well known that horizon area is a diffeomorphism invariant geometrical concept and is observer independent. Our derivation also provides credence to the understanding that symmetries, which preserves the boundary conditions give rise to global charges, which in this case is the Wald entropy/ modified geometric area. The local Lorentz symmetry may be viewed as a diffeomorphism for a Rindler like observer. Therefore, if one uses the metric variables, only those spacetime diffeomorphism  are useful which preserve the horizon. On the other hand, if one instead uses the tetrad and spin connection variables the horizon need to be invariant under local Lorentz transformations as well. These local Lorentz transformations are the ones which shall concern us here.\\

Since our whole analysis depends crucially on the notion of boundary, it is worth understanding how do we define a black hole boundary. Although it is clear that black hole boundary can not be a physical boundary as it does not represent the edge of the spacetime and an external observer can simply pass through this hypersurface. However, we can always impose certain mathematical conditions on this surface to ensure the presence of a one way membrane that resembles black hole horizon, and in this sense we call it an inner boundary. Such a boundary can be characterized by using the notion of event horizon (EH), Killing horizon (KH) or by using quasilocal frameworks like isolated horizon (IH) 
\cite{Ashtekar:1998sp,Ashtekar:1999ex,Ashtekar:1999yj,Ashtekar:1999wa,Ashtekar:2000sz,Ashtekar:2004cn,Ashtekar:2000hw,Ashtekar:1999sn,Ashtekar:2001is,Ashtekar:2003jh}.  During the last decade, it has been clearly understood that quasilocal definitions of black hole (and cosmological) horizons, like the Isolated (IH) and 
the Dynamical horizons (DH) formalisms
offer a much better description of the geometry and dynamics of horizons compared to those arising from the notion of event horizons (EH) since these definitions, in contrast to that of the EH, do not need to refer to the asymptotic infinities or the full spacetime to confirm its own existence. Indeed, the IH is defined quasilocally by imposing constraints on 
the local geometry in such a way that only the horizon is allowed to be stationary-the entire spacetime may contain matter and radiation with the requirement that they do not cross the horizon. Therefore, the IH formalism allows only the horizon to admit a timelike Killing vector: the exterior spacetime may be without any kind of equilibrium. Naturally, the IH formalism is an improvement over the EH formalism, which, being rectroactively defined, requires the knowledge of the future null infinity. Furthermore, the space of solutions belonging to the IH class is, therefore, quite large compared to those of the EH, since there exist solutions like those belonging to the Robinson- Trautman class \cite{Robinson:1960zzb} (which have spherical 
gravitational waves) for which the horizon is an IH but not an EH. The applications of these quasilocal definitions of horizon span vast area ranging from background independent quantum gravity to black hole mergers and numerical simulations related to physics of gravitational waves \cite{Ashtekar:2002ag,Dreyer:2002mx,Ashtekar:2025wnu}. These quasilocal definitions of black hole horizons have also opened up newer ways to critically appreciate the origin of Noetherian and Hamiltonian charges and their role in the classical and quantum mechanical understanding of black hole horizons.\\

The structure of the paper is outlined below: In the next section (\ref{2+1}), we will demonstrate our idea by calculating boost charge in $2+1$ dimensions. Although the calculation for $2+1$ dimensions has already appeared in \cite{Garg:2026mpo}. We nevertheless present it here for completeness and to make the discussion self-contained, since it will be useful to motivate subsequent analysis. We begin by specifying boundary conditions for IH in $2+1$ dimensions in section (\ref{3 bc}). The $2+1$ dimensional theory with negative cosmological constant admit a black hole solution known as BTZ black hole \cite{Banados:1992wn}, we show that BTZ horizon satisfy the boundary conditions. We then construct the full $SO(2,1)$ connection and its pullback to the horizon in (\ref{3 spin}). We show  that only a subgroup of full $SO(2,1)$ keeps the connection invariant under their action. In section (\ref{3 ps}) we briefly review the construction of symplectic structure and then compute Hamiltonian charge corresponding to boost and show that it is equal to the circumference of the horizon. In section (\ref{Lovelock}) we then apply the ideas to full Lovelock class of theories. We begin in section (\ref{4 bc}) by writing down the boundary conditions in arbitrary dimensions and then discuss how they differ from the boundary conditions in the  lower $[(2+1)~\text{or}~ (3+1)]$ dimensions. We then discuss the consequence of these boundary conditions on the geometry of the horizon and various geometric fields on the horizon in section (\ref{consequences}). In section (\ref{Riemann Weyl}) we discuss how these boundary conditions impose restrictions on Riemann and Weyl tensor. We then proceed to compute the spin connection in section (\ref{N spin}) and discuss the allowed Local Lorentz transformations that preserve the boundary conditions in section (\ref{N symmetries}). In section (\ref{N ps}) we construct the phase space and discuss the construction of the symplectic structure briefly and compute the Hamiltonian charge corresponding to the boost which turns out to be geometric area resembling the Wald entropy. And finally, in section (\ref{Discussions}) we discuss and summarize the findings and assumptions in our work.\\

\section{Local Lorentz charges for $2+1$ dimensions}\label{2+1}
The ideal place to begin our study of horizon charges is in the $2+1$ dimensions. Here, the horizon of a BTZ black hole provides the geometry to understand and extract thermodynamic quantities in a simplified setting. The isolated horizon formalism for $2+1$ dimensional spacetime was already set-up in \cite{Ashtekar:2002qc}. In the following, we shall briefly review the boundary conditions. As mentioned before, a discussion on the local Lorentz charges were carried out in great detail in \cite{Garg:2026mpo} in the context of developing a formalism 
to understand the Hawking radiation from a microscopic point of view. Our derivation here shall only contain the bare essentials, the details may be found in \cite{Garg:2026mpo}.

\subsection{Boundary Conditions}\label{3 bc}
Let $\mathcal{M}$  be a $3$-dimensional pseudo-Riemannian manifold endowed with metric $g_{ab}$ of signature $(-~+~+)$. A weak isolated horizon in the manifold $(\mathcal{M},g_{ab})$ is a null sub-manifold $\Delta$ generated by the congruence of  null normal $\ell^a$, equipped with a non-invertible metric $q_{ab}$ of signature $(0~+~+)$ which is naturally inherited from the spacetime metric such that the following conditions hold on $\Delta$: 
\begin{enumerate}
\item $\Delta$ is topologically $\mathbb{S}^{1}\times\mathbb{R}$ and null.
\item The expansion $\theta_{(\ell)}$ of the generator
vanishes on $\Delta$.
\item All the field equations holds true on $\Delta$, and $-T^{a}{}_{b}\ell^{b}$ is future directed and causal for all future directed $\ell^{a}$.
\item There exists a one form $\omega_{(\ell)}$, which correspond to the connection in normal bundle of $\mathbb{S}^{1}$,
is Lie-dragged by the generator $\ell^{a}$ on $\Delta$, that is $\lie_{\ell}\, \omega_{(\ell)}=0$. 
\end{enumerate}
The complete set of basis vectors contains the set ($\ell^{a}, n^{a}, m^{a}$), where $\ell^{a}$ and $n^{a}$ are null and $m^{a}$ is spacelike. In these basis, the metric, with signature (- + +) is given by $g_{ab}=-2\,\ell_{(a}\,n_{b)}+m_{a}\,m_{b}$. The non- vanishing dot products are $\ell^{a} n_{a}=-1, m^{a} m_{a}=1$. The Newman- Penrose (NP) scalars are obtained from the following expansion in terms of these basis set \cite{Ashtekar:2002qc}:
\begin{eqnarray}
\nabla_{a}\ell_{b}&=&-\epsilon\, n_{a}\ell_{b}+\kappa_{NP}\, n_{a} m_{b}-\gamma\, \ell_{a}\ell_{b}+\tau\, \ell_{a} m_{b}+\alpha\, m_{a}\ell_{b}-\rho\, m_{a}m_{b},\\
\nabla_{a}\, n_{b}&=&\epsilon\, n_{a}n_{b}-\pi\, n_{a}m_{b}+\gamma\, \ell_{a}n_{b}-\nu\, \ell_{a}\, m_{b}-\alpha\, m_{a}n_{b}+\mu\, m_{a}\, m_{b},\\
\nabla_{a}\, m_{b}&=&\kappa_{NP}\, n_{a}n_{b}-\pi\, n_{a}\ell_{b}+\tau\, \ell_{a}n_{b}-\nu\, \ell_{a}\ell_{b}-\rho\, m_{a}n_{b}+\mu\, m_{a}\ell_{b},
\end{eqnarray}
where the quantities $\epsilon, \rho, \gamma, \kappa_{NP}, \alpha, \mu, \nu$ and $\pi$ are the NP scalars. One may show that the well known BTZ solution given by the following metric, is an isolated horizon ($l^{2}\,\Lambda=-1$). 
\begin{eqnarray}
  &&  ds^{2}= -N^{2}\, dv^{2}+2dv dr + r^{2}\left[ d\phi +N^{\phi}\, dv\right],\\
&&N=\left[f(r)+\frac{J^{2}}{4\pi^{2}r^{2}}\right]^{1/2},~ N^{\phi}=\frac{-J}{2\pi r^{2}}, ~f(r)=-\frac{M}{\pi}+\frac{r^{2}}{\l^{2}}.
\end{eqnarray}
The BTZ solution is characterized by two parameters, namely $M$, and $J$. The scalar $N$ ($g_{vv}$) vanishes on the horizon and its roots gives us the locations of the horizon:
\begin{eqnarray}
r_{\mp}=\frac{M\l^{2}}{2\pi}\left[1\mp \left\{1-\left(\frac{J}{M\l}\right)^{2}\right\}^{1/2}\right]\, .
\end{eqnarray}
We denote the outer horizon by $r=r_{+}$. The full spacetime basis vectors for this solution are given by :
\begin{eqnarray}
\ell^{a}=\left(\frac{\partial}{\partial v}\right)^{a}+\frac{N^{2}}{2}(\partial_{r})^{a}-N^{\phi}\, (\partial_{\phi})^{a},~~~~
n^{a}=-(\partial_{r})^{a} ,~~~~m^{a}=\frac{1}{r}(\partial_{\phi})^{a}\, .
\end{eqnarray}
A simple calculation shows that the value of the Newman Pentose scalars are:
\begin{eqnarray}
\epsilon&=&[(1/2)f^{\prime}(r)- r(N^{\phi})^{2}], ~\pi=N^{\phi},~\kappa_{NP}=0, ~ \tau=N^{\phi} \nonumber\\
\rho&=&(-1/2r)N^{2}, ~\alpha=N^{\phi}, ~~ \nu=0, ~~ \mu=-(1/r),~~ \gamma=0.
\end{eqnarray}

Clearly, the outer horizon located at at $r=r_{+}$, has topology $\mathbb{S}^{1}\times \mathbb{R}$, since its generator is a Killing vector given by, 
 $\ell^{a}=(\partial_{v})^a-N^{\phi}(r_{+})(\partial_{\phi})^a$. Further the NP coefficient $\rho=0$  on the horizon which implies that the expansion scalar $\theta_{(\ell)}=-2\rho$ of the null generator is zero. Since BTZ is a vacuum solution the third condition is trivially true. So, the horizon at $r=r_{+}$,
 is an example of isolated horizon. \\
 
The laws of black hole mechanics are discussed in \cite{Ashtekar:2002qc} while a discussion on the mi. 
In the following, we shall only briefly delve into those issues, but keep our discussion mainly focused on derivation of the gauge charges on the IH. However, to keep the discussion complete, we shall derive the symplectic structure on the covariant phase- space of the $2+1$ gravity and obtain the Hamiltonian charges.\\

%
\subsection{Action, spin connection and symmetries} \label{3 spin}
As we have already discussed in the previous sections, 
our basic variables are the tetrad ($e_{a}^I$) and Lorentz Lie- algebra valued spin connection one-form (${A_{a}}^I$), where $I$ is the internal index that takes value in $SO(2,1)$ Lie algebra. The field strength $2$-form is given by $F_{I}=dA_{I}+(1/2)\,\epsilon_{IJK}\, A^{J}\wedge A^{K}$ through which we shall construct our first order 
Palatini action. The Lagrangian 3-form, with a cosmological constant $\Lambda$ takes the following form \cite{Romano:1991up}:
\begin{eqnarray}
8\pi G\, L=e^{I}\wedge F_{I} -(\Lambda/6)\,\epsilon^{IJK}\,e_{I}\wedge e_{J}\wedge_{K},
\end{eqnarray}
Variation with respect to $A^I$ will give:
\begin{equation}\label{de0}
    D e^{I}=de^{I}+A^{I}{}_{J}\wedge e^{J}=0 
\end{equation}
and the boundary term:
\begin{equation}\label{theta_sym}
    8\pi G~\theta(A,\delta A)= -e^I \wedge \delta A_I
\end{equation}
$\theta(\delta)$ is the symplectic potential. The above equation \eqref{de0} shows that the connection one- form $A_{I}$ is a spin- connection. A variation with respect to the tetrad variables give rise to the Einstein equations \cite{Ashtekar:2002qc}. The $2+1$ Palatini theory also admits an equivalent Chern-Simons theory with $ISO(2,1)$ as gauge group \cite{Witten:1988hc}, but for our 
analysis we shall use the Palatini theory. Note that the $\epsilon$ tensor allows a map, so that the connection $1$-form $A^{I}$ is given by $A_{IJ}=-\epsilon_{IJK}A^{K}$. This is the adjoint representation of the Lie algebra of $SO(2,1)$ with respect to the structure constants $\epsilon_{IJK}$.  Here, the connection $1$ form $A_{IJ}$ is a linear operator on the internal $3$- dimensional vector space equipped with the Minkowski metric $\eta_{IJ}$, satisfying $A_{IJ}=A_{[IJ]}$. The corresponding field strength $2$-form is given by $F_{IJ}=dA_{IJ}+[A, \, A]_{IJ}$, and $F^{I}=\epsilon^{IJK}\, F_{JK}$. Further, it is useful to express the epsilon tensor in terms of the NP basis: $\epsilon_{IJK}=3!\,\ell_{[I}n_{J}m_{K]}$, 
which gives,  $\epsilon^{IJK}\, m_{K}=\epsilon^{IJK}\, \epsilon_{K}=\epsilon_{IJ}=2\ell_{[I} n_{J]}$, and $(1/2)\epsilon^{IJK}\,\epsilon_{IJ}=-m_{K}=-\epsilon_{K}$. 
We can expand the spin connection in terms of internal basis $(\ell^I, n^I, m^I)$:
\begin{equation}
    A_{aI}=-\mathcal{\mathcal{W}}_{a} m_{I}-\mathcal{N}_{a}\, n_{I} +\mathcal{U}_{a}\, \ell_{I},
\end{equation}
where the coefficient one forms $\mathcal{W},\,\mathcal{N},  ~\text{and}~\mathcal{U}$ are given by:
\begin{eqnarray}
    \mathcal{W}=-\gamma \ell -\epsilon n+\alpha m \, ,~~~~
\mathcal{N}=\tau \ell +\kappa_{NP}\, n -\rho m \, , ~~~~
\mathcal{U}= -\nu \ell -\pi n +\mu m \, .
\end{eqnarray}
The full $SO(2,1)$ connection ${A_a}^{IJ}$  in the spacetime is given by:
\begin{eqnarray}
{A_{a}}^{IJ}&=& \mathcal{W}_a B^{IJ}+\mathcal{N}_a P^{IJ} +\mathcal{U}_a Q^{IJ}
\end{eqnarray}
where $B^{IJ}=-2\,\ell^{[I}n^{J]}$, $P^{IJ}=2\,n^{[I}m^{J]}$, and 
$Q^{IJ}=2\, \ell^{[I}m^{J]}.$ Local Lorentz transformations act on internal indices and transform the physical quantities as follows:
\begin{equation}\label{loc_lor}
e^{I}\rightarrow \Lambda^{I}{}_{J}\,e^{J}, ~~\text{whereas}~~~~ 
A^{IJ}\rightarrow (\Lambda^{-1})^{IK}\,{A_{K}}^{L}\,\Lambda_{L}{}^{J}+ (\Lambda^{-1})^{IK}\,d{\Lambda_{K}}^{J}.
\end{equation}
The local Lorentz transformations constitute the following series of transformations which preserve the NP basis:  
\begin{eqnarray}
\ell\rightarrow \eta \ell;\,~~ n\rightarrow \eta^{-1} n;\,~~ m \rightarrow m,\\
\ell\rightarrow \ell;~~\, n\rightarrow n-s m +(1/2)s^{2} \ell ;\,~~ m \rightarrow -s\ell +m,\\
\ell\rightarrow \ell -u m +(1/2)u^{2} n;\,~~ n\rightarrow n;\,~~ m \rightarrow -u n +m.
\end{eqnarray}
The connection components transform under these local Lorentz transformation in the following way:
\begin{eqnarray}
\mathcal{W}\rightarrow \mathcal{W}-d\ln\eta ;~ \mathcal{U}\rightarrow \eta \mathcal{U}; ~\mathcal{N}\rightarrow \eta^{-1} \mathcal{N}\\
\mathcal{W}\rightarrow \mathcal{W}-sN ;~ \mathcal{U}\rightarrow \mathcal{U}-ds+s\mathcal{W}-(1/2)s^{2}\mathcal{N};~ \mathcal{N}\rightarrow  \mathcal{N}\\
\mathcal{W}\rightarrow \mathcal{W}+u\mathcal{U} ;~ \mathcal{U}\rightarrow \mathcal{U};~ \mathcal{N}\rightarrow  \mathcal{N} +du-u\mathcal{W}-(1/2)u^{2}\mathcal{U}.
\end{eqnarray}
For a black hole horizon $\mathcal{N}\stackrel{\Delta}{=}0$ in any basis, and hence, the third transformation is not useful since $\mathcal{N}$ does not transform homogeneously. The reason for this non-homogeneous transformation of $\mathcal{N}$ can be traced back to the non-homogeneous transformation of $\kappa_{NP}$ and $\rho$ which are related to the geodeticity and expansion of the horizon generator $\ell^a$. On the other hand, the first two, a null boost in the $\ell- n$ plane, and a null rotation in $\ell-m$ plane needs to be studied as symmetries. In 4-dimensions the residual symmetry group on the horizon is $ISO(2)\ltimes \mathbb{R}$ \cite{Basu:2010hv}. \\

\subsection{Phase space and Hamiltonian charge} \label{3 ps}
For the phase space analysis we start with the
symplectic potential $\theta(\delta)$, as derived in eqn. \eqref{theta_sym}. the quantity $\theta(A,\delta A)$ is a $2$-form on the spacetime, but it can be viewed as a $1$-form on the phase space. One can construct from the pre-symplectic potential a symplectic current $2$-form $J$, by antisymmetrizing the arbitrary second variation of pre-symplectic $2$-form (on configrarion space this is equivalent to taking exterior derivative of $\theta$):
\begin{equation}
    J(\delta_1,\delta_2)=\delta_1\theta(A,\delta_2 A)-\delta_2\theta(A,\delta_1 A).
\end{equation}
The symplectic current can be written as: 
\begin{equation}\label{symplectic_current 1}
    8\pi GJ(\delta_1,\delta_2)=-(\delta_1e^I\wedge \delta_2A^I-\delta_1e^I\wedge \delta_2A^I)
\end{equation}
Let us now construct the symplectic structure:  
\begin{figure}\label{2+1 IH}
    \centering
    \includegraphics[width=0.48\linewidth]{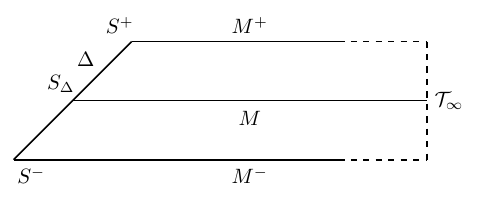}
    \caption{The figure shows the spacetime manifold $\mathcal{M}$ bounded by past and future partial Cauchy slices $M^{\pm}$ that intersect $\Delta$ in the 2-spheres $S^{\pm}$ respectively, and extend to spatial infinity which is a timelike cylinder  $\tau_\infty$. Another Cauchy slice $M$ which intersects $\Delta$ in $S_{\Delta}$ is drawn.}
    \label{fig1}
\end{figure}
Let us consider the manifold $\mathcal{M}$ as shown in fig (\ref{fig1}). Since $J(\delta_{1}, \delta_{2})$ is closed, on $\mathcal{M}=\Delta\cup M^{+}\cup M^{-}\cup \tau_{\infty}$, we get:
\begin{eqnarray}\label{symp}
\left[\, \int_{M^{+}} -\int_{M^{-}} -\int_{\Delta} \,\right] \, J(\delta_{1}, \delta_{2})= 0,
\end{eqnarray}
where the contribution of symplectic current on $\tau_{\infty}$ can be dropped out with a suitable choice of fall-off conditions on the fields. On $\Sigma^{+}$ or $\Sigma^{-}$, the value of $J(\delta_{1}, \delta_{2})$ is as given in \eqref{symplectic_current 1}, but on $\Delta$, the expression become an exact form such that $J(\delta_1,\delta_2)=dj(\delta_1,\delta_2)$ and hence we can write the hypersurface independent symplectic structure as:
\begin{equation}
    8 \pi G \, \Omega(\delta_1,\delta_2)=\int_{M} J(\delta_1,\delta_2)-\int_{S_{\Delta}} j(\delta_1,\delta_2)
\end{equation}
From here we get \cite{Ashtekar:2002qc}:
\begin{eqnarray}\label{symplectic1}
8\pi G\, \Omega(\delta_{1}, \delta_{2})=-\int_{M}\left[\delta_{1}A^{I}\,\wedge\delta_{2}\, e_{I}-\delta_{2}A^{I}\, \wedge \delta_{1}\, e_{I} \right] -\oint_{S_{\Delta}}\left[\delta_{1}\psi\, \delta_{2} m -\delta_{2}\psi\, \delta_{1} m \right].
\end{eqnarray}
where $\psi$ is the potential corresponding to the surface gravity $\kappa_{(\ell)}$ and is defined as $\pounds_\ell \psi\stackrel{\Delta}{=}\kappa$ such that $\psi|_{S_-}=0$. 
We can now proceed with the calculation of the Hamiltonian charge corresponding to local Lorentz boost.
For infinitesimal local Lorentz transformations $
    \lambda_{I}{}^{J}=(\delta_{I}{}^{J} +\epsilon_{I}{}^{J})$, one gets the transformations of triads and connection one form:
\begin{eqnarray}
\delta_{\lambda} e^{I}&=&\epsilon^{I}{}_{J}\, e^{J}, \label{tetrad and connection 1}\\
\delta_\lambda A^{IJ}&=&-d\lambda^{IJ} -A^{IK}\lambda_{K}{}^{J}-A^{JK}\,\lambda^{I}{}_{K}.\label{tetrad and connection 2}
\end{eqnarray}
A simple calculation gives that for the transformation $\delta_{\lambda}$ a boost, with $\epsilon_{IJ}=-2\ell_{[I}\, n_{J]}$, the Hamiltonian charge is the area (circumference) of the cross- section,  
\begin{equation}
\Omega(\delta_{\lambda}, \delta)= -\oint_{S_{\Delta}}\, \epsilon_{IJK}\delta e^{K} \times\epsilon^{IJ}=\oint_{S_{\Delta}}\delta\, m =\delta\, (2\pi r_{+}).
\end{equation}
%
%
%
%
%

\section{Local Lorentz Charges For Lovelock Theory In Arbitrary Dimensions }\label{Lovelock}
In the following, we shall extend the formalism of the previous section for higher dimensional spacetimes and for a general class of diffeomorphism invariant class of action. We shall first define a non- expanding horizon for higher dimensions, and derive the consequence of these boundary conditions on the horizon geometry. We shall thereafter construct the phase space and calculate the Hamiltonian charge for the horizon boost.
\subsection{Boundary Conditions} \label{4 bc}
Let $(\mathcal{M}, g_{ab}, \nabla)$ be a $N$- dimensional spacetime equipped with metric $g_{ab}$ of signature 
$(-++...)$, and a  covariant derivative operator $\nabla$ compatible with the metric. Let $\Delta$ be a null hypersurface generated by the integral curves of future directed null normal $\ell^{a}$, equipped with a non-invertible metric inherited naturally from the spacetime metric. Hereafter, we shall use the symbol 
$\stackrel{\Delta}{=}$ to denote equalities that hold true only on the horizon, and an arrow under an index $T_{\underleftarrow{a}}$ is used to denote the pullback of that quantity to the horizon. The induced metric on $\Delta$ is  given by $q_{ab}{\stackrel{\Delta}{=}}g_{\underleftarrow{ab}}$. $\Delta$ is said to be an weak isolated horizon (WIH) if the following conditions hold:
\begin{enumerate}
    \item $\Delta$ is topologically $\mathbb{S}\times \mathbb{R}$, where $\mathbb{S}$ is an $(N-2)$ dimensional compact and connected surface.
    \item The congruence of the null normal $\ell^{a}$ is expansion free, that is, $\theta_{(\ell)}{\stackrel{\Delta}{=}}0$, where $\theta_{(\ell)}=q^{ab}\, \nabla_{a}\ell_{b}$.
    \item $-R_{a}{}^{b}\,\ell^{a}$ is future directed null vector field, and all field equations hold on $\Delta$.
    \item The quantity $[\lie_{\ell}, \, \nabla_{\underleftarrow{a}}]\ell^a\,{\stackrel{\Delta}{=}}\,0$.
\end{enumerate}
In the first requirement, we keep the topology of horizon cross sections to be sufficiently general to include possible geometries that a horizon can have in higher dimensions \cite{Galloway:2005mf}. 
The second condition implements the 
expansion freeness of the horizon, which in turn implies that the horizon remain null throughout. This also implies that no matter falls through the horizon during the period of observation. Note that this is the most crucial information on the geometry of the horizon. The third condition is equivalent to dominant energy condition (usually put on stress energy tensor), is directly placed on Ricci tensor. This is because in higher dimensions, field equations may take complicated forms and that makes it difficult to relate stress energy tensor to Ricci tensor. Although, it is shown in \cite{Devdutt:2026bmlt} that for spherically symmetric WIH, the condition on the Ricci tensor can be translated directly into a condition on stress energy tensor. As we shall see below, the fourth condition is equivalent to demanding that connection in the normal bundle $\omega_{(\ell)}$ is Lie dragged by horizon generators $\ell^{a}$. Such a condition (along with expansion freeness) is similar to the requirement that both the induced metric and the extrinsic curvature are Lie dragged by the horizon generator). As was pointed out in \cite{Ashtekar:2000hw} this condition is sufficient to derive the laws of black hole mechanics in general relativity. 

\subsection{Geometry of the horizon and consequences of the boundary conditions} \label{consequences}
First, we set up the basis vector fields adapted to the horizon. Let the horizon $\Delta$ be foliated 
by codimension-2 spacelike sub-manifolds $S_{v}$ labeled by $v=$constant. To span the space normal to the spheres $S_{v}$, we choose two null vector fields  $\ell^{a}$ and $n^{a}$. A good natural choice for the covector $n_{a}$ is $n_{a}=-(dv)_{a}$, so that $\ell^{a}=\partial_{v}^{a}$ is tangential to the horizon $\Delta$. The spacelike vectors which span the  tangent space $T(S_v)$ of the spatial cross sections are denoted by $m_{(i)}^a$. Note that since the spatial indices ($i,j,k...$) are raised or lowered by $\delta_{ij}$, there is no difference between covariant and contravariant components labeled by ($i,j,k...$). To construct Newman-Penrose like formalism let us introduce following notation which we have adopted from \cite{Coley:2004jv,Pravda:2004ka,bib23,Ortaggio:2009bz,Ortaggio:2009zt,Ortaggio:2012cp,Ortaggio:2012jd,Ortaggio:2014sea,Chatterjee:2025jur,Devdutt:2026bmlt}
\begin{align}\label{ortho1}
    m^{(0)}&=n & m^{(1)}&=\ell & m^{(i)}&= \text{spacelike covectors}\\ \label{ortho2}
    m_{(0)}&=-\ell & m_{(1)}&=-n &  m_{(i)}&= \text{spacelike vectors},
\end{align}
These vector fields obey following orthonormality conditions: $\ell^{a} n_{a}=-1,\, \ell^{a} m_{a}^{(i)}=0,\, n^{a} m_{a}^{(i)}=0,\, m^{a(i) }m_{a}^{(j)}=\delta^{(ij)}$. In line with the proposals of \cite{Coley:2004jv,Pravda:2004ka,bib23,Ortaggio:2009bz,Ortaggio:2009zt,Ortaggio:2012cp,Ortaggio:2012jd,Ortaggio:2014sea}, we use the following notation: a common label $(a)$ in $m^{(a)}$ is introduced to denote all the frame vectors, \emph{i.e.} the indices $a,b,c, \cdots$ run from $0$ to $(N-1)$ whereas a new set of  indices $i,j,k, \cdots$ that run from $2$ to $(N-1)$, is used to represent only spacelike vectors $m^{(i)}$. This NP- like basis allows us to write the spacetime metric as:
\begin{equation}
    g_{ab}=-2\ell_{(a}n_{b)}+\delta_{ij} m^{(i)}_a m^{(j)}_b
\end{equation}
The derivatives of various frame vectors can be written in terms of Ricci rotation coefficients defined as follows:
\begin{eqnarray}\label{Ricci_coeff}
    \nabla_{b} \,\ell_{a} &=& L_{cd}\, m_a^{(c)}m_b^{(d)}\label{100}\\
    \nabla_{b} \, n_{a}& =& N_{cd}\, m_a^{(c)}m_b^{(d)}\\
    \nabla_{b} \, m_{a}^{(i)}& =&  \stackrel{(i)}{M}_{cd}\, m_a^{(c)}m_b^{(d)}
\end{eqnarray}
Notably, all of these Ricci rotation coefficients are not independent. There are several constraints which are imposed on them, first due to orthonormality conditions and second, due to the boundary conditions. Let us determine the constraints  on the Ricci rotation coefficients which arise due to the orthonormality conditions on the vectors in \eqref{ortho1}, and $\eqref{ortho2}$:
\begin{eqnarray}\label{Ricci_cond}
   L_{i a} - \stackrel{i}{M_{0a}}&=0, ~~  & N_{0a}+L_{1a}=0, ~~~ L_{0a} = 0,\\\label{Ricci_cond_1}
    N_{i a} - \stackrel{i}{M_{1a}}&=0, ~~~ & \stackrel{j}{M_{ia}}+\stackrel{i}{M_{ja}} =0,~~~  \stackrel{i}{M_{ia}} =0, ~~~   N_{1a}=0. \label{Ricci_cond_2}
\end{eqnarray}
Next, we use the boundary conditions for the WIH, as given in the previous subsection to further constrain the Ricci coefficients. First, since $\ell^{a}$ is a null normal it has to be a geodetic vector field, such that
\begin{equation}
\ell^{b}\nabla_{b}\, \ell^{a}\propto \ell^{a}.
\end{equation}
As a consequence, it is easy to show that, $L_{i0}\stackrel{\Delta}{=}0$, which we can obtain as follows: Expand equation \eqref{Ricci_coeff} using the constraints in equation \eqref{Ricci_cond_1} to obtain: 
\begin{equation}
\ell^b\nabla_b \ell_{a}=L_{c0} m_a^{(c)}=L_{10}\ell^{a}+L_{i0}m^{(i)}_{a}.
\end{equation}
But since the vector field $\ell^a$ is geodetic, one requires that the right side be proportional to $\ell^{a}$, which implies that $L_{i0}\stackrel{\Delta}{=}0$. Using $L_{0a}=0$ and $L_{i0}=0$ we can write:
\begin{equation}\label{grad bl}
    \nabla_{\underleftarrow{b}}\ell^{a} \stackrel{\Delta}{=}\left[L_{10} n_b + L_{1i} m_b^{(i)}\right] \,\ell^{a} + L_{ij}m_b^{(j)}m^{(i)\,a}.
\end{equation}
The quantity in the square brackets above is the connection on the normal bundle, and is given by:
\begin{equation}\label{omegasmall}
    \omega_b^{(\ell)}=L_{10} n_b + L_{1i} m_b^{(i)}.
\end{equation}
One also calls this $\omega^{(\ell)}$ as the rotation one-form and is expresed as:
\begin{equation}\label{omega_expansion}
        \omega_{\underleftarrow{a}}^{(\ell)}\stackrel{\Delta}{=}-\kappa_{(\ell)} n_a + \Tilde{\omega}_{a}^{(\ell)},
\end{equation}
where $\Tilde{\omega}^{(\ell)}$ is the pullback of $\omega^{(\ell)}$ onto the spatial cross sections of the horizon, which usually vanish for spherically symmetric black holes. \\

Next, we can decompose the matrix $L_{ij}$ in its symmetric (and tracefree) and the antisymmetric parts as follows: $L_{ij}=S_{ij}+A_{ij}$, where these quantities are given by:
\begin{equation}
S_{ij}=L_{(ij)}=\sigma^{(\ell)}_{ij}+\theta_{(\ell)}\delta_{ij}\, , ~~~~ A_{ij}=L_{[ij]}=\Omega^{(\ell)}_{ij}\, ,
\end{equation}
where $\Omega^{(\ell)}_{ij}$ is the twist corresponding to the generator $\ell^{a}$.  Since $\ell^a$ is null normal it must be twist free, $\Omega^{(\ell)}_{ij}\stackrel{\Delta}{=}0$. Next, to obtain further constraint, let us consider the Raychaudhuri equation:
\begin{equation}\label{Raychaudhuri}
\pounds_{\ell}\,\theta_{(\ell)}\stackrel{\Delta}{=}\kappa_{(\ell)}\theta_{(\ell)}-\frac{1}{(N-2)}\,\theta_{(\ell)}^{2} +|\Omega^{(\ell)}_{ij}|^{2}- |\sigma^{(\ell)}_{ij}|^{2} -R_{ab}\,\ell^{a}\,\ell^{b} \, .
\end{equation}
Since the WIH $\Delta$ is expansion- free and twist free, it follows that:
\begin{equation}\label{Raychaudhuri_1}
  |\sigma^{(\ell)}_{ij}|^{2} + R_{ab}\,\ell^{a}\,\ell^{b} \, \stackrel{\Delta}{=}0.
\end{equation}
Furthermore, the boundary condition on the Ricci tensor on a given WIH (the third condition on WIH) and the above equation \eqref{Raychaudhuri_1} gives that: 
\begin{eqnarray}\label{IH Ricci}
     \sigma^{(\ell)}_{ij}\stackrel{\Delta}{=}0,~~\text{and } ~~R_{ab}\ell^a \ell^b\stackrel{\Delta}{=}0.
 \end{eqnarray}
Therefore, black hole boundary, here denoted by a WIH, is generated by vector field $\ell^{a}$ which is twist-free, shear-free and expansion-free and consequently $L_{ij}\stackrel{\Delta}{=}0$. The vanishing of these Newman–Penrose scalars is the consequence of WIH boundary conditions. As we shall see, the following two conditions on $ L_{i0}$, and $L_{ij}$ will play crucial role in deciding which Lorentz transformations survive on $\Delta$ to form a residual symmetry group.
\begin{equation}\label{BC}
    L_{i0}\stackrel{\Delta}{=}L_{ij}\stackrel{\Delta}{=}0
\end{equation}
To summarise the present calculation until now, a consequence of these developments 
along with \eqref{grad bl} and \eqref{omegasmall} gives:
\begin{equation}
    \nabla_{\underleftarrow{b}}\ell^a \stackrel{\Delta}{=}\omega_b^{(\ell)}\ell^a
\end{equation}
Furthermore, one can use eqns. \eqref{Ricci_cond}, and eqns. \eqref{Ricci_cond_1}, to show that the following Ricci rotation coefficients also vanish identically on $\Delta$:
\begin{equation}
    L_{0a}\stackrel{\Delta}{=}L_{i0}\stackrel{\Delta}{=}L_{ij}\stackrel{\Delta}{=}\,\stackrel{i}{M_{00}}\,\stackrel{\Delta}{=}\,\stackrel{i}{M_{0j}}\,\stackrel{\Delta}{=}0. 
\end{equation}
Our choice of $n=-dv$ which implies that $\underleftarrow{dn}=0$, gives rise to more constraints on the Ricci rotation coefficients:
\begin{align}\label{n=dv}
     L_{1i}\stackrel{\Delta}{=}-N_{i0}, ~~~N_{[ij]}\stackrel{\Delta}{=}0.
\end{align}
A simple calculation of the action of the horizon generator $\ell^{a}$ on the vector fields tangential to the horizon cross- section, along with equations \eqref{n=dv}  gives:
\begin{equation}\label{lie m}
     \lie_\ell m^a_{(i)}\stackrel{\Delta}{=}-(N_{i0}+L_{1i})\ell^{a}-\stackrel{i}{M_{j0}} m_{(j)}^a
    m^{a}_{(i)}\stackrel{\Delta}{=}-\stackrel{i}{M_{j0}} m_{(j)}^{a}\, .
\end{equation}
%

\subsection{Conditions on the Riemann tensor and the Weyl tensor} \label{Riemann Weyl}
In the previous section, as a consequence of boundary condition we found that $R_{ab}\ell^a\ell^b\stackrel{\Delta}{=}0.$ 
This condition when translated in terms of stress energy tensor implies that their is no flux of matter across the horizon \cite{Devdutt:2026bmlt}. The above equation implies that $-{R_{a}}^b\l^a$ is tangent to $\Delta$ and can not have any component in the direction of $n^a$. And then from third boundary condition we know that $-{R_{a}}^b\l^a$ is a future directed causal vector field. Therefore, 
$-{R_{a}}^b\l^a$ can only be proportional to $\ell^a$, 
and consequently we have the following:
\begin{eqnarray}\label{NP_matterfield_values}
R_{\underleftarrow{a}b}\,\ell^{b}\, \stackrel{\Delta}{=} 0\, , ~~~~ R_{ab}\l^{a}\l^{b}\, \stackrel{\Delta}{=} 0\, ~~~~ R_{ab}\l^{a}m^{b}_{(i)}\stackrel{\Delta}{=}\,0\, .
\end{eqnarray}
We can use  $2\nabla_{[a}\nabla_{b]}\ell^{c}=R_{ab}{}^{c}{}_{d}\, \ell^{d}$, to show that:
\begin{equation}\label{Riemann tensor condition}
   \tensor{R}{_{\underleftarrow{ab}}^{c}_{d}}\,\ell^{d}\,\stackrel{\Delta}{=} [d\omega^{(\ell)}]_{\underleftarrow{ab}}\,
   \ell^{c}\, .
\end{equation}
Since $R_{\underleftarrow{a}b}\ell^b\stackrel{\Delta}{=}0$, we have following conditions (for the definition of these components see appendix (\ref{Appendix A})):
\begin{align}
    R_{0i0i}\stackrel{\Delta}{=}0,~~~R_{0kik}+R_{010i}\stackrel{\Delta}{=}0
\end{align}
Further from \eqref{Riemann tensor condition} we get:
\begin{align}
    R_{0i0j}\stackrel{\Delta}{=}R_{0ijk}\stackrel{\Delta}{=}0
\end{align}
Now we can find out the restrictions imposed by isolated horizon boundary conditions on various Weyl scalars on the horizon and then establish the relation between the components of Weyl tensor and rotation one form $\omega$. 
First, note that the Weyl tensor and the 
curvature tensors are related by the following equation:
\begin{equation}
    C_{abcd}=R_{abcd}-\frac{2}{N-2}(g_{a[c]}R_{d]b}-\, g_{b[c]}R_{d]a})+\frac{2}{(N-1)(N-2)}R\,g_{a[c}\,g_{d]b}.
\end{equation}
First, from $R_{\underleftarrow{a}b}\ell^{b}\stackrel{\Delta}{=}0$, and the fact that $\ell_{\underleftarrow{a}}\stackrel{\Delta}{=}0$, it follows that the following equation holds on $\Delta$:
\begin{equation}\label{Weyl Riemann}
   { R_{\underleftarrow{ab} d}}^{c}\, \ell^{d}\stackrel{\Delta}{=} { C_{\underleftarrow{ab} d}}^{c}\, \ell^{d}
\end{equation}
And then we can use the equation \eqref{Riemann tensor condition} to show that the exterior derivative of the rotation one form $\omega^{(\ell)}$ is given by:
\begin{equation}\label{Weyl omega}
[\,d{\omega^{(\ell)}}\, ]_{\underleftarrow{ab}}\stackrel{\Delta}{=} C_{\underleftarrow{ab}cd}\,\ell^{c} n^{d}\, . 
\end{equation}
Using the equation \eqref{Riemann tensor condition} and the equation \eqref{Weyl Riemann} it can be shown without much effort that:
\begin{align}
    C_{0i0j}\stackrel{\Delta}{=}C_{0ijk}\stackrel{\Delta}{=}0
\end{align}
Further using the traceless property of the Weyl tensor \emph{i.e.} $g^{ac}C_{abcd}=0$ and the fact that $C_{0ijk}\stackrel{\Delta}{=}0$, it can be shown that:
\begin{equation}
    C_{010i}\stackrel{\Delta}{=}0
\end{equation}
From the expansion of the Weyl tensor in terms of the basis vector one forms (see \eqref{Weyl expansion} in appendix (\ref{Appendix A})) one gets that:
\begin{equation}
    C_{abcd}\,\ell^c n^d= C_{010i}[n\wedge m^{(i)}]_{ab}+C_{0101}(n\wedge \ell)_{ab} +\frac{1}{2} C_{01ij}\, [\, m^{(i)} \wedge m^{(j)}\,]_{ab}-C_{101i}\, \,[\,\ell \wedge m^{(i)}\,]_{ab}\, .
\end{equation}
Therefore, on the horizon, the expansion results in a much more compact form:
\begin{equation}
C_{\underleftarrow{ab}cd}\,\ell^c n^d\stackrel{\Delta}{=} C_{010i}\, [\,n\wedge m^{(i)}\,]_{\underleftarrow{ab}} +\frac{1}{2} C_{01ij}\, \,[\,m^{(i)} \wedge m^{(j)}\, ]_{\underleftarrow{ab}}\, .
\end{equation}
 But the Weyl component $C_{010i}\stackrel{\Delta}{=}0$. Let us define $C_{ij}:=(1/2)\,C_{01ij}$, and then we have:
\begin{equation}\label{Weyl on horizon}
C_{\underleftarrow{ab}cd}\,\ell^c n^d\stackrel{\Delta}{=}\frac{1}{2} C_{01ij}\,[m^{(i)} \wedge m^{(j)}]_{\underleftarrow{ab}}\stackrel{\Delta}{=}\, C_{ij}\,[m^{(i)} \wedge m^{(j)}]_{\underleftarrow{ab}}\, .
\end{equation}
From equations \eqref{Weyl omega} and \eqref{Weyl on horizon} we get:
\begin{equation}\label{spatial domega}
[d{\omega^{(\ell)}}]_{\underleftarrow{ab}}\stackrel{\Delta}{=}\, C_{ij}\, [m^{(i)} \wedge m^{(j)}]_{\, \underleftarrow{ab}}.
\end{equation}
The equation \eqref{spatial domega} implies that the projection of $d\omega^{(\ell)}$ on the $\ell-n$ plane is zero and therefore it is a purely spatial $2$-form. This result is crucial as it plays an important role for the construction of symplectic structure. From the fourth boundary condition, $[\lie_{\ell}, \, \nabla_{\underleftarrow{a}}]\ell^a\,{\stackrel{\Delta}{=}}\,0$ we find that $d\lie_{\ell}\, \omega\stackrel{\Delta}{=}0$. Since the tangential vectors $m^{(i)}$ of the spherical cross- sections are lie dragged by $\ell^{a}$, 
and that $\lie_{\ell}\, d\omega^{(\ell)}\stackrel{\Delta}{=}\,0$, we get from the previous equation \eqref{spatial domega} that $\lie_\ell C_{ij}\stackrel{\Delta}{=}0$. In four dimensions $C_{ij}$ are related to the imaginary part of the Weyl scalar $\Psi_{2}$, and one has that $d\omega= \text{Im}\Psi_{2}\, {}^{2}\epsilon$, where the quantity ${}^{2}\epsilon$ is the volume form on the 2-sphere cross- section. Since the volume form on the cross-section of the horizon in 4-dimensions is Lie-dragged along the horizon generator $\ell^a$, one has that the $\operatorname{Im}\Psi_2$ is also Lie dragged along $\ell^a$. But this does not hold true in higher dimensions, since there is no canonical choice of a $2$- form on the horizon cross- section which may be preferred over any other. However, the Weyl scalars remain the same as one moves along the horizon generator $\ell^a$. In higher dimensions their evolution along the horizon generator is tied to the choice of vector fields that span tangent space of spatial cross sections. With an appropriate choice of these vector fields on the horizon, however, one can ensure that the Weyl scalars do not change along $\ell^a$, but that will put constraints on the $\stackrel{i}{M}_{j0}$.
To see this, note that on the horizon:
\begin{equation}
\lie_{\ell} m^{(i)}_{\underleftarrow{b}}\stackrel{\Delta}{=}-\stackrel{i}{M_{j0}}m^{(j)}_{\underleftarrow{b}}.
\end{equation}
So if we put $\lie_\ell C_{ij}\stackrel{\Delta}{=}0$, this will give rise to following constraint:
\begin{equation}\label{constraint 1}
   \begin{array}{c @{\quad\text{or}\quad} c}
 C_{ik} \stackrel{k}{M_{j0}}-\stackrel{i}{M_{k0}} C_{kj}\stackrel{\Delta}{=}0 & [C,M_0]\stackrel{\Delta}{=}0
\end{array}
\end{equation}
One can easily verify that the above constraint is trivially satisfied in $4$-dimensions as expected. One gets another constraint as follows: if $\phi^{a}$ is the rotational symmetry vector field on the horizon, then one can expect that the connection on the normal bundle is also Lie
dragged along this vector field, \emph{i.e.} $\lie_\phi \, \omega \stackrel{\Delta}{=} 0$. We also assume that 
the intrinsic geometry of the cross-sections do not alter along the horizon generator, since no matter falls through the horizon during our observation time. This implies that
the components of the Riemann tensor on the cross- sections are lie- dragged: $\lie_{\ell} \,\mathcal{R}_{ijkl}\stackrel{\Delta}{=}0$, and we get following constraint:
\begin{equation}\label{Riemann constraint}
( \mathcal{R}_{n[j|kl}\stackrel{n}{M}_{|i]0}+\mathcal{R}_{ijn[l}\stackrel{n}{M}_{k]0})\stackrel{\Delta}{=}0\, .
\end{equation}
It can be verified that $\lie_\ell m^a_{(i)}\stackrel{\Delta}{=}0$ is one choice that satisfies both the constraint of equation \eqref{constraint 1} and equation \eqref{Riemann constraint}. But once we make this choice for the basis vector fields on the spatial cross sections, we are restricted to make arbitrary spin transformations of $m^{a}_{(i)}$ only on the initial cross section $S_-$ . In the subsequent analysis we will stick to this choice of basis vector fields on spatial cross sections that satisfy $\lie_\ell m^a_{(i)}\stackrel{\Delta}{=}0$.

\subsection{Spin connection $A_{IJ}$ and its field strength $F_{IJ}$ on the horizon} \label{N spin}
For our next set of calculations, we shall need the spacetime connection. The full $SO(N-1,1)$ connection on the spacetime can be written in terms of the local Lorentz basis as:
\begin{equation}\label{connection_expansion}
    \tensor{A}{_{a}^{IJ}}=-2 \mathcal{U}_{a}\, \ell^{[I}n^{J]}+2 \mathcal{N}^{(i)}_{a}\, n^{[I}m^{J]}_{(i)}+2 \mathcal{V}^{(i)}_{a} \, \ell^{[I}m^{J]}_{(i)}+ \mathcal{W}^{(ij)}_{a}\, m_{(i)}^{[I}m^{J]}_{(j)}.
\end{equation}
But again as a consequence of boundary conditions one can show that $\mathcal{N}_a^{(i)}\stackrel{\Delta}{=}0$. More precisely, $\mathcal{N}_{a}^{(i)}$ contains Newman-Penrose coefficients that vanish on the horizon due to the horizon generator $\ell^a$ being geodetic, expansion free, and shear free. It may be argued that $A_{IJ}$ on the horizon takes the following form:
\begin{equation}\label{connection_expansion}
    \tensor{A}{_{\underleftarrow{a}}^{IJ}}\stackrel{\Delta}{=}-2 \omega^{(\ell)}_{a}\, \ell^{[I}n^{J]}+2 \mathcal{V}^{(i)}_{a}\, \ell^{[I}m^{J]}_{(i)}+ \mathcal{W}^{(ij)}_{a}\, m_{(i)}^{[I}m^{J]}_{(j)}.
\end{equation}
To obtain this we proceed as follows: Let the
internal basis $[\,\ell^{I}, n^{I}, m^{(i)\, I}\,]$ are fixed on the horizon such that 
they are annihilated by the derivative operator $\partial_{a}$, and hence, $\nabla_{a}\ell^{I}=A_{a}^{I}{}_{\,J}\, \ell^{J}$. Now, using 
the equation \eqref{BC} to rewrite the equation $\nabla_{\underleftarrow{a}}\ell^{b}\stackrel{\Delta}{=}\omega_{a} \ell^b$, and from these two different expressions:
\begin{equation}
    \nabla_{\underleftarrow{a}} \ell^{b}\stackrel{\Delta}{=} \omega^{(\ell)}_{a} \ell^{b} \stackrel{\Delta}{=} \nabla_{\underleftarrow{a}}(e^{b}_I \,\ell^I)=e^{b}_{I}\, {A_{\underleftarrow{a}}}^{I}{}_{J}\ell^{J},
\end{equation}
we find that the component of spin-connection in $\ell^I-n^I$ plane should take the form: $A^{IJ}=-\omega_{a}\, \ell^{I}\, n^{J}$.
Similarly the action of the covariant derivatives on remaining  vector fields can also be expressed in terms of coefficient one forms:
\begin{eqnarray}\label{11}
    \nabla_{\underleftarrow{a}}n^b&\stackrel{\Delta}{=} &-\omega^{(\ell)}_{a} n^{b} +\mathcal{V}^{(i)}_a m_{(i)}^{b}\\
\nabla_{\underleftarrow{a}}m^b_{(i)}&\stackrel{\Delta}{=}&\mathcal{V}_a^{(i)}\ell^b-\mathcal{W}_a^{(ij)}m^b_{(j)}, ~~~~~~~~~\, (i\ne j).
\end{eqnarray}
Finally, repeating the similar calculation as we did to find the projection of spin connection in $\ell^I-n^I$ plane, we can find out the other coefficient one-forms as well:
\begin{eqnarray}\label{W and V}
\mathcal{V}^{(i)}_{\underleftarrow{b}}&\stackrel{\Delta}{=}& N_{i0}n_b+N_{ij}m^{(j)}_{b}\\
\mathcal{W}^{ij}_{\underleftarrow{b}}&\stackrel{\Delta}{=}&-\stackrel{i}{M_{j0}}n_b -\stackrel{i}{M_{jk}}m^{(k)}_b
\end{eqnarray}
Once we have calculated the spin connection in terms of internal basis naturally the the next thing is to compute the field strength or curvature $F_{IJ}$ corresponding to the spin connection $A_{IJ}$ in equation 
\eqref{connection_expansion}. The curvature of spin connection is given by $F_{IJ}=dA_{IJ}+A_{IK}\wedge\, A^{K}{}_{J}$. A straightforward calculation reveals that:
\begin{equation}
    {F_{\underleftarrow{c d}}}^{IJ}\stackrel{\Delta}{=}-2 (d\omega^{(\ell)})_{cd} \, l^{[I}n^{J]}+2 [d\mathcal{V}^{(k)}+\{\omega \wedge \mathcal{V}^{(k)}\}+\{\mathcal{V}^{(i)}\wedge \mathcal{W}^{(ik)}\}]_{cd}\,  l^{[I}m^{J]}_{(k)} +[d\mathcal{W}^{(ik)}+ \{\mathcal{W}^{(ij)} \wedge \mathcal{W}^{(jk)}\}]_{cd}\, m^{I}_{(i)} m^{J}_{(k)}.
\end{equation}
We can use following relation along with  equation \eqref{Riemann tensor condition} to write $F^{IJ}$ in a form which is useful in subsequent analysis:
\begin{equation} \label{dW}
    d\mathcal{W}^{(ik)}\stackrel{\Delta}{=}-\mathcal{W}^{(ij)}\wedge \mathcal{W}^{(jk)} + \mathcal{R}_{abpq} m^{p}_{(i)} m^{q}_{(k)}
\end{equation}
The above equation \eqref{dW} can be obtained 
directly from the definition of the field strength corresponding to connection $1$- form  $\mathcal{W}^{(ij)}$ on the spatial cross section of $\Delta$. The quantity $\mathcal{R}_{cdps}$ is the Riemann tensor associated with the metric on the horizon cross-sections.
\begin{equation}\label{curvature}
     {F_{\underleftarrow{c d}}}^{IJ}\stackrel{\Delta}{=} -2 (d{\omega^{(\ell)}})_{cd} \, l^{[I}n^{J]}+2 [d\mathcal{V}^{(k)} +\{\omega^{(\ell)} \wedge \mathcal{V}^{(k)}\}+\{\mathcal{V}^{(i)}\wedge \mathcal{W}^{(ik)}\}]_{cd}\,  l^{[I}m^{J]}_{(k)} + \tensor{\mathcal{R}}{_{kl}^{ij}} m_c^{(k)}m_d^{(l)}m^I_{(i)}m^J_{(j)}
\end{equation}
For our subsequent analysis of the horizon symmetries, the action and the symplectic structure plays an important role, and the above expression of the field strength $F_{IJ}$ is to play an important role.
\subsection{Local Lorentz symmetries on the Isolated horizon in $N$- dimensions} \label{N symmetries}
Once a set of basis vector fields is chosen on the horizon, it is always possible to apply local Lorentz transformations to obtain an alternative basis \cite{Chandrasekhar:1985kt,Carmeli:1977,Milson:2004jx}. On $\Delta$ the transformations belong to four different classes: ${(i)}$ boost in $\ell-n$ plane, ${(ii)}$ spatial rotations or spins $(iii)$ null rotations which leave $\ell$ unchanged, and $(iv)$ null rotations which leave $n$ unchanged.
\begin{align}
         &(i)~~\ell\xrightarrow{}\eta \ell,~~n\xrightarrow{} \eta^{-1} n,~~m^{(i)}\xrightarrow{}m^{(i)}\\
        &(ii)~~\ell\xrightarrow{} \ell,~~n\xrightarrow{}  n,~~m^{(i)}\xrightarrow{} {R^i}_j m^{(j)}\\
        &(iii)~~\ell\xrightarrow{} \ell,~~ n\xrightarrow{} n+ \mathcal{F}_{i}m^{(i)}+\frac{1}{2}\mathcal{F}^2 \ell,~~m^{(i)}\xrightarrow{} m^{(i)}+\mathcal{F}_i \ell\\
        &(iv)~~ \ell\xrightarrow{}\ell+\mathcal{F}_i m^{(i)}+\frac{1}{2}\mathcal{F}^2 n,~~n\xrightarrow{}n,~~ m^{(i)}\xrightarrow{}m^{(i)}+\mathcal{F}_i n
\end{align}
    In $(ii)$ ${R^i}_j$ are $(3\times3)$ orthogonal matrices. In $(iii)$ and $(iv)$ $\mathcal{F}_i$ are real functions such that $\mathcal{F}_i \mathcal{F}^i=\mathcal{F}^2$. Recall from the boundary conditions, of equation \eqref{BC} that,  $L_{i0}\stackrel{\Delta}{=}L_{ij}\stackrel{\Delta}{=}0$. We want to keep only those transformations that preserve these boundary conditions, \emph{i.e.} we will allow only those transformations under which these Ricci coefficients transform homogeneously. This shall ensure that the transformations map $\Delta\rightarrow\Delta$. Under the first three transformations these Ricci coefficients transform homogeneously as shown below \cite{bib23}:
     \begin{align}
         &(i)~~L_{i0}\xrightarrow{}\eta^2 L_{i0},~~L_{ij}\xrightarrow{} \eta L_{ij},\\
        &(ii)~~ L_{i0}\xrightarrow{}{R^i}_j  L_{j0},~~L_{ij}\xrightarrow{}  {R^i}_k {R^j}_l L_{kl},\\
        &(iii)~~L_{i0}\xrightarrow{} L_{i0},~~ L_{ij}\xrightarrow{} L_{ij}-\mathcal{F}_jL_{j0} .
    \end{align}
However, these Ricci rotation coefficients do not transform homogeneously under the fourth transformations, as shown below:
\begin{multline}
    L_{i0}\xrightarrow{}L_{i0}-\mathcal{F}_j L_{ij}+\mathcal{F}_i N_{00}+\mathcal{F}_j \stackrel{j}{M}_{i0}+\frac{1}{2}\mathcal{F}^2(L_{i1}+N_{i0})-\mathcal{F}_i\mathcal{F}_j(N_{0j}+N_{j0})-\mathcal{F}_j\mathcal{F}_k \stackrel{j}{M}_{ik}+\mathcal{F}_i\mathcal{F}_j\mathcal{F}_k N_{jk}\\
    +\frac{1}{2}\mathcal{F}^2(\mathcal{F}_i N_{01}-\mathcal{F}_j N_{ij}+\mathcal{F}_j \stackrel{j}{M}_{i1})+\frac{1}{2}\mathcal{F}^2(-\mathcal{F}_i\mathcal{F}_j N_{j1}+\frac{1}{2}\mathcal{F}^2 N_{i1})-\ell^a\nabla_a \mathcal{F}_i-\mathcal{F}_j m^a_{(j)}\nabla_a \mathcal{F}_i-\frac{1}{2}\mathcal{F}^2n^a\nabla_a \mathcal{F}_i
\end{multline}
\begin{multline}
    L_{ij}\xrightarrow{} L_{ij}+\mathcal{F}_i N_{0j} -\mathcal{F}_jL_{i1}+\mathcal{F}_k \stackrel{k}{M}_{ij}-\mathcal{F}_j(\mathcal{F}_i N_{01}+\mathcal{F}_k \stackrel{k}{M}_{i1})-\mathcal{F}_i\mathcal{F}_k N_{kj}+\frac{1}{2}\mathcal{F}^2 N_{ij}+\mathcal{F}_i\mathcal{F}_j\mathcal{F}_kN_{k1}\\
    -\frac{1}{2}\mathcal{F}^2 \mathcal{F}_j N_{i1}+m_{(j)}^a\nabla_a \mathcal{F}_i+\mathcal{F}_j n^a\nabla_a \mathcal{F}_i
\end{multline}
Consequently, the coefficient one-form $N^{(i)}_a$ in the spin connection will transform inhomogeneously under the transformation $(iv)$. This implies that the transformation $(iv)$ does not preserve the boundary conditions on the horizon. This is not surprising, as the rotation about vector field $n^a$ can transform $\ell^a$ to make it point off the horizon, and hence does not keep the horizon invariant. We can also prove that the transformation that keeps $n$ invariant does not preserve the boundary conditions by analyzing the transformation properties of connection one form $A^{IJ}$.  Under the local Lorentz transformations the connection one transform as follows:
\begin{equation}
    A_{IJ}\xrightarrow{} {\Lambda_I}^{L}\, A_{LK} \, \Lambda^{K}{}_{J}+\Lambda_{IL}\, d{\Lambda_J}^{L}
\end{equation}
Here $\Lambda_{IJ}$ are Lorentz matrices associated with the local Lorentz transformations and are given by:
\begin{align}
{(i)}\qquad
\Lambda_{IJ}
&=-\eta\,\ell_I n_J-\eta^{-1}n_I\ell_J
+\delta_{ij}m^{(i)}_I m^{(j)}_J,\\
{(ii)}\qquad
\Lambda_{IJ}
&=-\ell_I n_J-n_I\ell_J
+R^i{}_j\,m^{(j)}_{I}m^{(i)}_{J},\\
{(iii)}\qquad
\Lambda_{IJ}
&=-\ell_I n_J
-\left(n_I+\mathcal F_i m^{(i)}_{I}
+\frac12\mathcal F^2\ell_I\right)\ell_J
+\left(m^{(i)}_{I}+\mathcal F_i\ell_I\right)m^{(i)}_{J},\\
{(iv)}\qquad
\Lambda_{IJ}
&=-\left(\ell_I+\mathcal F_i m^{(i)}_{I}
+\frac12\mathcal F^2n_I\right)n_J
-n_I\ell_J
+\left(m^{(i)}_{I}+\mathcal F_i n_I\right)m^{(i)}_{J}.
\end{align}
Under above transformations $(i)-(iv)$, respectively, the connection
one-forms transform as:
\begin{gather}
\mathcal U_a\xrightarrow{}\mathcal U_a-\partial_a\ln \eta,\quad
\mathcal N_a^{(i)}\xrightarrow{}\eta^{-1}\mathcal N_a^{(i)},\quad
\mathcal V_a^{(i)}\xrightarrow{}\eta\mathcal V_a^{(i)},\quad
\mathcal W_a^{(ij)}\xrightarrow{}\mathcal W_a^{(ij)}.
\\[5pt]
\begin{aligned}
&\mathcal U_a\xrightarrow{}\mathcal U_a,\quad
\mathcal N_a^{(i)}\xrightarrow{}R^j{}_i\mathcal N_a^{(j)},\quad\mathcal W_a^{(ij)}
\xrightarrow{}R^k{}_iR^l{}_j\mathcal W_a^{(kl)}
-(\partial_aR^k{}_i)R^k{}_j, \quad
\mathcal V_a^{(i)}\xrightarrow{}R^j{}_i\mathcal V_a^{(j)}.
\end{aligned}
\\[5pt]
\begin{aligned}
&\mathcal U_a\xrightarrow{}\mathcal U_a+\mathcal F_i\mathcal N_a^{(i)},\quad
\mathcal N_a^{(i)}\xrightarrow{}\mathcal N_a^{(i)},\quad
\mathcal W_a^{(ij)}\xrightarrow{}\mathcal W_a^{(ij)}
+\mathcal F_i\mathcal N_a^{(j)}-\mathcal F_j\mathcal N_a^{(i)},\\
&\mathcal V_a^{(i)}
\xrightarrow{}\mathcal V_a^{(i)}-\partial_a\mathcal F_i
-\mathcal F_i\mathcal U_a-\mathcal W_a^{(ij)}\mathcal F_j
+\frac12\mathcal F^2\mathcal N_a^{(i)}
-\mathcal F_i\mathcal F_j\mathcal N_a^{(j)}.
\end{aligned}
\\[5pt]
\begin{aligned}
&\mathcal U_a\xrightarrow{}\mathcal U_a-\mathcal F_i\mathcal V_a^{(i)},\quad
\mathcal V_a^{(i)}\xrightarrow{}\mathcal V_a^{(i)},\quad
\mathcal W_a^{(ij)}\xrightarrow{}\mathcal W_a^{(ij)}
+\mathcal F_i\mathcal V_a^{(j)}-\mathcal F_j\mathcal V_a^{(i)},\\
&\mathcal N_a^{(i)}
\xrightarrow{}\mathcal N_a^{(i)}-\partial_a\mathcal F_i
+\mathcal F_i\mathcal U_a-\mathcal W_a^{(ij)}\mathcal F_j
+\frac12\mathcal F^2\mathcal V_a^{(i)}
-\mathcal F_i\mathcal F_j\mathcal V_a^{(j)}.
\end{aligned}
\end{gather}
Under the first three transformations, the connection one form $\mathcal{N}^{(i)}$ transforms homogeneously. So, if it vanishes on the horizon in one frame, it will also vanish in the transformed one. However, under the fourth transformation $\mathcal{N}^{(i)}$ does not transform homogeneously. Therefore, the fourth transformation is not allowed. In the following, we want to compute the Hamiltonian charge corresponding to the local Lorentz boost transformation in the $\ell-n$ plane. The generator of the local Lorentz boost symmetry is given by:
$\epsilon_{IJ}=-2\,l_{[I}\, n_{J]}$.
 
\subsection{Phase space and Hamiltonian charge} \label{N ps}
 Our analysis of phase- space and the Hamiltonian charges 
 shall be carried out on an $N$ dimensional space time $\mathcal{M}$ which is bounded by the inner boundary $\Delta$, two  Cauchy slices $M_1$ and $M_2$ in future and past respectively and the spatial asymptotic infinity $i^{0}$ (see figure \ref{fig2}). The action contains the following dynamical variables: Lie algebra valued tetrad $N$-bein fields $e^I_a$ and $SO(N-1,1)$ Lie algebra valued connection one-form $A^{IJ}$.
The Lagrangian $N$-form for the Lovelock theory on a $N$ dimensional manifold $\mathcal{M}$ is given by the following \cite{Lovelock:1971yv,Zanelli:2005sa, Zumino:1985dp,Banados:1993ur}:
\begin{figure}\label{manifold}
    \centering
    \includegraphics[width=0.50\linewidth]{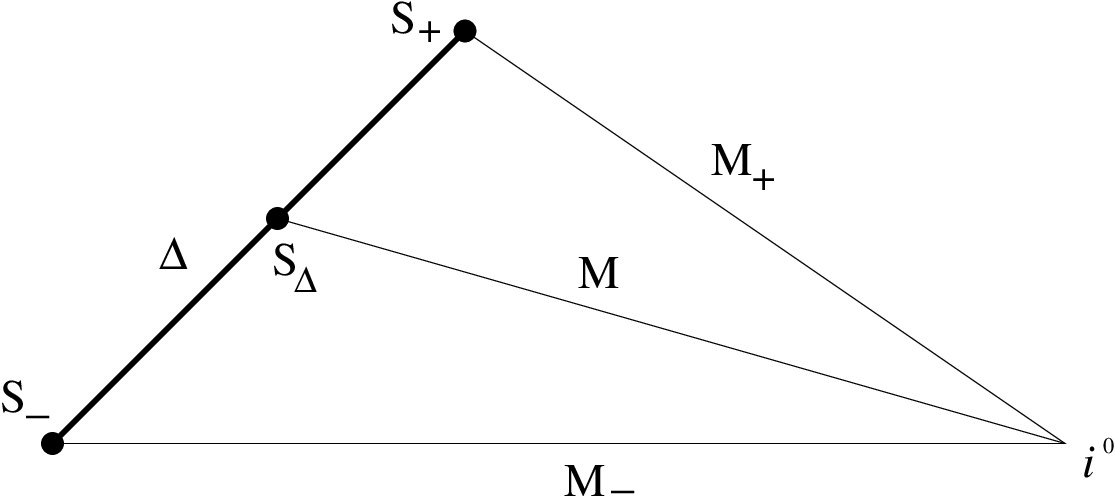}
    \caption{The figure shows the spacetime manifold $\mathcal{M}$ bounded by past and future partial Cauchy slices $M_{\pm}$ that intersect $\Delta$ in the 2-spheres $S_{\pm}$ respectively, and extend to spatial infinity $i^{0}$. Another Cauchy slice $M$ which intersects $\Delta$ in $S_{\Delta}$ is drawn.}
    \label{fig2}
\end{figure}
%
\begin{equation}\label{Lagrangian}
   16 \pi G \, L\,[\,e, \, A]= \sum_{p=0}^{[N/2]} \alpha_{p}\, L^{(p)},
\end{equation}
where $[\cdot]$ denotes the floor function and the coefficients $\alpha_p$ are arbitrary constants. The quantity $L^{(p)}$ is given by the following expression:
\begin{equation}
    L^{(p)}=\frac{1}{(N-2p)!}\epsilon_{a_1...a_{2p} a_{2p+1}...a_{N}} F^{a_1 a_2} \wedge...\wedge F^{a_{2p-1} a_{2p}}\wedge e^{a_{2p+1}} \wedge...\wedge e^{a_{N}},
\end{equation}
and $F_{IJ}=dA_{IJ}+
A_{IK}\wedge A^{K}{}_{J}$ is the field strength corresponding to 
the connection variable one form $A_{IJ}$. Before proceeding, it is important to highlight several features of the above Lagrangian: First of all, the Lagrangian on a $N$-dimensional manifold is a polynomial in a curvature 2-form with degree $[N/2]$, but has same number of degrees of freedom as the second order metric variable based action in $N$-dimensions\footnote{By maximizing the number of degrees of freedom in the theory, it has been shown that for odd dimensions the theory can be written as Chern-Simons theory for (A)dS or Poincare group and in even dimensions it can be written as Born-Infeld theory \cite{Troncoso:1999pk}.}.
Second, the $p=0$ term does not get any contribution from the curvature $F_{IJ}$ and hence its variation with respect to the connection is always zero. Similarly when $N$ is even, the $p = [N/2]$ term does not get any contribution from $e^{I}_{a}$ and hence it also has a trivial variation with respect to the $N$- beins. Moreover, in even dimensions the variation of $p = [N/2]$ term with respect to spin connection does not contribute to equation of motion because of the Bianchi identity, $DF^{IJ}=0$. Therefore, $p = [N/2]$ term is topological and is an Euler invariant on an even $N$-dimensional manifold. Consequently, we can even truncate our series in action eqn. \eqref{Lagrangian} to $p=[(N-1)/2]$. We restrict ourselves to the vanishing cosmological constant case and therefore the series in action eqn. \eqref{Lagrangian} starts from $p=1$. Even though this topological term does not contribute to the equations of motion, it can give rise to a boundary term that may contribute to the symplectic potential but leaves the symplectic current invariant (a careful discussion of these issues has been carried out in a series of papers \cite{Corichi:2010ur,Corichi:2013zza,Corichi:2016zac,Corichi:2016eoe}). The variation with respect to the  $e^{I}_{a}$ does not give rise to any boundary term and yields the following gravitational equations of motion:
\begin{equation}\label{tetrad eom}
    \sum_{p=1}^{\left[(N-1)/2\right]}  \frac{1}{(N-2p-1)!} \alpha_{p}\, \mathcal{E}^{p}_{I_{N}}=0
\end{equation}
where,
\begin{equation}
\mathcal{E}^p_{I_{N}}=\epsilon_{I_1...I_{2p} I_{2p+1}...I_{N}}\, F^{I_1 I_2} \wedge...\wedge F^{I_{2p-1} I_{2p}}\wedge e^{I_{2p+1}} \wedge...\wedge e^{I_{N-1}}.
\end{equation}
On the other hand, if we vary the action with respect to the connection $A^{IJ}$ we get the following equation of motion:
\begin{equation}\label{connection eom}
    D \Tilde{\Sigma}_{I J}=(-1)^{N+1}\,d \Tilde{\Sigma}_{IJ} +\Tilde{\Sigma}_{IK}\wedge {A^K}_J + \Tilde{\Sigma}_{KJ}\wedge {A^K}_I=0
\end{equation}
 where the operator $D$ is the gauge covariant derivative, $D\psi^{I}=\partial\psi^{I}+A^{I}{}_{J}\, \psi^{J}$, for any internal vector $\psi^{I}$. Also 
\begin{equation}\label{Sigma 1}
    \Tilde{\Sigma}_{IJ}:= \sum_{p=1}^{\left[\frac{N-1}{2}\right]} p\alpha_p\, \Sigma_{IJ}^{p}
\end{equation}
where, the quantities of $\Sigma_{IJ} ^{p}$ are given by:
\begin{equation}\label{Sigma p}
    \Sigma_{IJ}^{p}=\frac{1}{(N-2p)!}\,\epsilon_{I J I_3 I_4...I_{2p} I_{2p+1}...I_{N}}\, F^{I_3 I_4}\wedge...\wedge F^{I_{2p-1} I_{2p}} \wedge e^{I_{2p+1}}\wedge...\wedge e^{I_N}
\end{equation}
The basis $e^{I}_{a}$ may be expanded in terms of the Lorentz basis on the full spacetime as follows:
\begin{equation}\label{tetrad}
   e^{I}_{a}=-n_a \ell^I-\ell_a n^I +m_a^{(i)}m_{(i)}^I
\end{equation}
Using equations \eqref{tetrad} and \eqref{curvature} we may write the  $\Tilde{\Sigma}_{IJ}$ on horizon in terms of internal basis vectors fields:
\begin{equation}\label{Sigmacompact}
    \Tilde{\Sigma}_{IJ}=2~{}^{N-2}\epsilon~ \mathcal{L} ~\ell_{[I}n_{J]}~+2~{}^{N-2}\epsilon~ \Tilde{C}_{k l}~ \, m_{I}^{(k)} m_{J}^{(l)}+ 2 \alpha_{(k)}\, \ell_{[I}m_{J]}^{(k)},
\end{equation}
where $\alpha_{k}$ in the above expression is some $N-2$ form, ${}^{N-2}\epsilon~$ is the area form on the spherical sections of the horizon, while the quantities $\mathcal{L}$ and $\Tilde{C}_{kl}$ are defined as follows:
\begin{equation}
    \mathcal{L}:=\sum_{p=1}^{\bigl[\frac{N-1}{2}\bigr]} p \alpha_p\left(\frac{1}{2^{p-1}}\right)
\delta^{i_1 i_2 \cdots i_{2p-3} i_{2p-2}}_{j_1 j_2 \cdots j_{2p-3} j_{2p-2}}
{\mathcal{R}_{i_1 i_2}}^{j_1 j_2}\cdots
{\mathcal{R}_{i_{2p-3}\, i_{2p-2}}}^{j_{2p-3} j_{2p-2}},
\end{equation}
\begin{equation}
    \Tilde{C}_{kl}:=\sum_{p=1}^{\left[\frac{N-1}{2}\right]} p\alpha_p
\left(\frac{p-1}{2^{p-2}}\right)
\delta^{ij\,i_1 i_2 \cdots i_{2p-5}\, i_{2p-4}}_{kl\,j_1 j_2 \cdots j_{2p-5}\,j_{2p-4}}
{\mathcal{R}_{i_1 i_2}}^{j_1 j_2}\cdots
{\mathcal{R}_{i_{2p-5}\, i_{2p-4}}}^{j_{2p-5} j_{2p-4}}
\,C_{ij}.
\end{equation}
$\mathcal{L}$ is known as the Lanczos-Lovelock scalar on the horizon. \\

Let us now look ate the variation of the action.
When the Euler term is not included in the Lagrangian 
the variation of the action on shell can be expressed as:
\begin{equation}
 \delta S=   (-1)^N \int_{\partial M=\Delta\,\cup\, i^{0}} \,\Tilde{\Sigma}_{IJ} \wedge  \delta A^{IJ}.
\end{equation}
 This needs to vanish for the differentiability of the action. Once again we can take care of terms which are at the asymptotic boundary by imposing suitable fall-off conditions on the dynamical fields at $i^{0}$, but the terms on the horizon must be handled carefully. We show that the boundary conditions force this integral to vanishes on $\Delta$. Using the expressions of spin-connection from equation \eqref{connection_expansion},  the corresponding curvature from equation \eqref{curvature}, and $\Tilde{\Sigma}_{IJ}$ from equation \eqref{Sigmacompact}, we get
\begin{equation}\label{Integrand}
 \widetilde{\Sigma}_{IJ} \wedge \delta A^{IJ}
\stackrel{\Delta}{=} 2~ \mathcal{L}~^{N-2} \epsilon\wedge \delta \omega^{(\ell)}+2 ~\Tilde{C}_{kl}~^{N-2} \epsilon \wedge \delta \mathcal{W}^{(kl)}
\end{equation}
Let us look at the first term in the above expression: The field $\omega$ is kept fixed on $S_{-}$. Since $\delta \ell=c\ell$ where $c$ is some constant, the WIH condition implies that $\lie_{\ell} \,\omega^{(\ell)} \stackrel{\Delta}{=} 0$. 
As a consequence, $\lie_{\ell} \,\,\delta \omega^{(\ell)}\stackrel{\Delta}{=}0$, and hence $\delta \omega^{(\ell)} \stackrel{\Delta}{=}0$. So first term vanishes on $\Delta$. The second term vanishes because of the gauge choice $\lie_\ell m^{(i)} \stackrel{\Delta}{=}0$. As a consequence, the surface term on $\Delta$ is zero due to the boundary conditions and the action is differentiable, and hence equations of motion can be extracted from $\delta S=0$.\\

 The symplectic structure can be obtained as follows: Notice that in its most general form, the variation of the Lagrangian $N$-form  can be expressed as: $\delta L=\mathcal{E}_{a} \delta \phi^{a}  +d \theta(\phi,\delta \phi)$, where $\phi$ is a common label for dynamical fields and $\mathcal{E}$ represents the field equations. The quantity $\theta(\phi,\delta \phi)$ being the symplectic potential, which can be thought of a spacetime $(N-1)$ form, and phase space $1$- form. One can construct from the pre-symplectic potential a symplectic current $2$-form $J$, by antisymmetrizing the arbitrary second variation of pre-symplectic $2$-form \cite{ASHTEKAR1991417, Wald:1993nt}:
\begin{equation}
    J(\delta_1,\delta_2)=\delta_1\theta(A,\delta_2 A)-\delta_2\theta(A,\delta_1 A).
\end{equation}
One can identify the form of symplectic potential from the Lagrangian in \eqref{Lagrangian}: $\theta(\delta)=\Tilde{\Sigma}_{IJ}\wedge \delta A^{IJ}$. The symplectic current can be written as:
 \begin{equation}\label{symplectic_current}
16 \pi G \,J(\delta_{1}, \delta_{2})=(-1)^N (\delta_1 \Tilde{\Sigma}_{IJ} \wedge \delta_2 A^{IJ} - \delta_2\Tilde{\Sigma}_{IJ} \wedge \delta_1 A^{IJ}).
\end{equation}
To get conserved symplectic structure from symplectic current consider a manifold $\mathcal{M}$ as shown in fig (\ref{fig2}). Since $J(\delta_{1}, \delta_{2})$ is closed, on $\mathcal{M}=\Delta\cup M^{+}\cup M^{-}\cup i^{0}$, we get:
\begin{eqnarray}\label{symp}
\left[\, \int_{M^+} -\int_{M^{-}} -\int_{\Delta} \,\right] \, J(\delta_{1}, \delta_{2})= 0,
\end{eqnarray}
where the quantity on $i^{0}$ can be dropped out by imposing suitable fall- off conditions on the fields. The expression of  $J(\delta_{1}, \delta_{2})$ on $M_{+}$ or $M_{-}$, is given in \eqref{symplectic_current}, however, on $\Delta$ one expects that $J$ can at most survive as a total derivative, since we are dealing with equilibrium situations and there is no matter flux across $\Delta$. Let us examine the form of $J$ on $\Delta$:
\begin{equation}
(-1)^{N}\, \,16 \pi G\, J(\delta_1, \delta_2)|_{\Delta}
\stackrel{\Delta}{=} 2\, \delta_1 (~^{N-2}\epsilon~\mathcal{L}) \wedge \delta_{2} \omega^{(\ell)} + 2 \delta_1(~^{N-2}\epsilon ~\Tilde{C}_{kl})\wedge \delta_2 \mathcal{W}^{(kl)}-(\delta_1 \leftrightarrow\delta_2)\, .
\end{equation}
Using the equations \eqref{omega_expansion} and 
the equation \eqref{W and V}, the above expression can take
the following form:
\begin{equation}\label{sympcurr_1}
(-1)^{N}\,16 \pi G\, J(\delta_1, \delta_2)|_{\Delta}
\stackrel{\Delta}{=} 2 \delta_1 (~^{N-2}\epsilon~\mathcal{L}) \wedge \delta_2 (-\kappa_{(\ell)} n) + 2 \delta_1(~^{N-2}\epsilon ~\Tilde{C}_{kl})\wedge \delta_2 (-\stackrel{k}{M_{l0}}n)-(\delta_1 \leftrightarrow\delta_2)
\end{equation}
The second term in the above expression vanishes again due to the gauge choice $\lie_\ell m^{(i)}\stackrel{\Delta}{=}0$. Now we define following potential for $\kappa_{\ell}$:
\begin{equation}\label{potential}
 \lie_\ell\, \psi\stackrel{\Delta}{=}\kappa_{(l)}=\omega^{(\ell)}\cdot\ell\, ,
\end{equation}
such that $\psi$ vanishes on $S_{-}$. Such a condition implies that on $\Delta$, we must have $d\psi==-\kappa_{(\ell)}\, n$. Using this expression into the equation \eqref{sympcurr_1}, we get the following
simplified form:
\begin{equation}
16 \pi G\, J(\delta_1, \delta_2)|_{\Delta}
\stackrel{\Delta}{=}2~ d[\delta_1(\mathcal{L}~^{N-2} \epsilon)\, \delta_{2}\, \psi-(\delta_1\leftrightarrow \delta_2)].
\end{equation}
The above result shows that the symplectic current reduces to an exact form on the horizon such that $J(\delta_1,\delta_2)|_{\Delta}=dj(\delta_1,\delta_2)$ and hence it only makes contribution on inital and final spatial cross sections $S_{\pm}$. It follows from \eqref{symp} the conserved symplectic structure can be written as:
\begin{equation}
    16 \pi G \, \Omega(\delta_1,\delta_2)=\int_M J(\delta_1,\delta_2)-\int_{S_{\Delta}} j(\delta_1,\delta_2)
\end{equation}
 Therefore, the hypersurface independent symplectic structure includes contribution both from the bulk and the horizon sections, and is given by:
\begin{equation}\label{symplectic structure}
16 \pi G\, \Omega(\delta_1, \delta_2)
=(-1)^N \int_{M} \, \left[\,\delta_{1} \Tilde{\Sigma}_{IJ}\wedge \delta_{2} A^{IJ}-\delta_{2}\Tilde{\Sigma}_{IJ}\wedge \delta_{1} A^{IJ}\,\right]-2\oint_{S_\Delta}\, \left[\delta_{1}(\mathcal{L}~^{N-2} \epsilon)\,\delta_{2} \psi-\delta_{2}(\mathcal{L}~^{N-2} \epsilon)\,\delta_{1}\, \psi\, \right].
\end{equation}
\\

Given this symplectic structure, let us determine the Hamiltonian charges corresponding to the local Lorentz transformations. If we donote the Lorentz transformations as the action of $\delta_{\lambda}$, then a function on the phase space $H_{\lambda}$ is the Hamiltonian corresponding to that phase space vector field if \cite{ASHTEKAR1991417, Wald:1993nt}:
\begin{equation}
    \Omega(\delta, \delta_{\lambda})=\delta H_{\lambda}.
\end{equation}

 Using the action of infinitesimal Lorentz transformations on the tetrad and connection variables \eqref{tetrad and connection 1} and \eqref{tetrad and connection 2}, the bulk contribution to the symplectic structure \eqref{symplectic structure} can be expressed in the following form (for the proof see appendix (\ref{A2})):
\begin{equation}\label{bulk symplectic structure}
\int_M (-1)^N\, (\delta \Tilde{\Sigma}_{IJ} \wedge \delta_\lambda A^{IJ} - \delta_\lambda\Tilde{\Sigma}_{IJ} \wedge \delta A^{IJ})=-\int_M d(\epsilon^{IJ} \delta \Tilde{\Sigma} _{IJ})=-\oint_{S_{\Delta}} \epsilon^{IJ} \delta \Tilde{\Sigma} _{IJ}
\end{equation}
Now let us compute the contribution of the boundary part of the symplectic structure to Hamiltonian charge. From \eqref{potential} we get:
\begin{equation} \label{psi variation}
    d \psi_{(\ell)}= -\kappa_{(\ell)}n + \alpha_{(\ell)}^{(i)} m_{(i)}
\end{equation}
On horizon $(\ell^{I}, n^{I}, m^{I}_{(i)})$ all are fixed. $\epsilon^{IJ}= \ell^{[I}n^{J]}$. $\delta_\lambda n_{a}=\delta_\lambda e^{I}_{a}\, n_{I}= ({\epsilon^I}_{J}\, e^{J}_{a})\, n_{I}=n_{a}.$ Similarly we can find out the action of infinitesimal Lorentz boost on other spacetime co-vectors.
\begin{eqnarray}\label{vector variation}
    \delta_\lambda n_a=n_a,~~\delta_\lambda \ell_a=-\ell_a,~~\delta_\lambda m_a^{(i)}=0 
\end{eqnarray}
 Now consider the equation: $\ell^{a}\nabla_{a} \ell^{b}=\kappa_{(\ell)} \ell^{b}$. If we use $\delta_\lambda$ on both sides of the equation:
 \begin{eqnarray}
     \delta_\lambda (\ell^a\nabla_a \ell^b)&\stackrel{\Delta}{=}&\delta_{\lambda}(\kappa_{\ell}\ell^b)\\ \delta_\lambda(\ell^a)\nabla_a\ell^b+\ell^a\delta_\lambda(\nabla_a \ell^ b)&\stackrel{\Delta}{=}&\delta_\lambda(\kappa_{\ell})\ell^b+\kappa_{\ell}\delta_\lambda \ell^b\label{kappa variation}
 \end{eqnarray}
 Since the spacetime metric remains invariant under the action of infinitesimal local Lorentz boost we can infer, $[\delta_\lambda, \nabla_a]=0$. Using this equation and the equation \eqref{vector variation} in equation \eqref{kappa variation}, we can show that
 $\delta_\lambda(\kappa_{(\ell)})= -\kappa_{(l)}$. Now we can use these results in \eqref{psi variation} to find out that $\delta_\lambda (d\psi)=\delta_\lambda [\,\alpha_{(\ell)}^{(i)}\,]\, m_{(i)}$ which gives $\ell \cdot d\, [\,\delta_\lambda \psi_{(\ell)}\, ]=\lie_{\ell}\, [\delta_\lambda \psi_{(l)}\, ]=0$. Hence $(\delta_\lambda \psi_{(l)})$ depends only on the coordinates of the spatial cross sections of $\Delta$. If we set $(\delta_\lambda \psi_{(l)})=0$  at some initial cross section, it vanishes over $\Delta$. Since $\delta_\lambda m^{(i)}_a=0$ we have $\delta_\lambda (^3 \epsilon)=0$. Finally, $\delta_\lambda (\mathcal{R}_{ijkl})=0$,  because the local Lorentz boost does not change the spacetime  metric, and consequently $\delta_\lambda \mathcal{L}=0$. Hence, we have proved that the boundary part of the symplectic structure does not make any contribution to the Hamiltonian charge corresponding to the local Lorentz boost. And finally, from \eqref{Sigmacompact},\eqref{symplectic structure} and \eqref{bulk symplectic structure} we find that:
\begin{equation}
    16 \pi G \,\Omega(\delta, \delta_\lambda)= \delta\oint_{S_\Delta}{}^{N-2}\epsilon~ \mathcal{L}=\delta\biggl[\oint_{S_{\Delta}} \,^{N-2}\epsilon\,\sum_{p=1}^{\left[\frac{N-1}{2}\right]} p\alpha_p\left(
\frac{1}{2^{\,p-1}}
\right)
\delta^{i_1 i_2 \cdots i_{2p-3}\, i_{2p-2}}_{j_1 j_2 \cdots j_{2p-3}\, j_{2p-2}}
\,{\mathcal{R}_{i_1 i_2}}^{j_1 j_2}\cdots
{\mathcal{R}_{i_{2p-3}\, i_{2p-2}}}^{j_{2p-3} j_{2p-2}}
\Biggr] 
\end{equation}
This is  a total variation, and hence the Hamiltonian function corresponding to the local Lorentz boost is given by: 
\begin{equation}\label{boost charge}
    H_\lambda=\oint_{S_\Delta}{}^{N-2}\epsilon~ \mathcal{L}=\oint_{S_{\Delta}} \,^{N-2}\epsilon\biggl[\,\sum_{p=1}^{\left[\frac{N-1}{2}\right]} p\alpha_p\left(
\frac{1}{2^{\,p-1}}
\right)
\delta^{i_1 i_2 \cdots i_{2p-3}\, i_{2p-2}}_{j_1 j_2 \cdots j_{2p-3}\, j_{2p-2}}
\,{\mathcal{R}_{i_1 i_2}}^{j_1 j_2}\cdots
{\mathcal{R}_{i_{2p-3}\, i_{2p-2}}}^{j_{2p-3} j_{2p-2}}
\Biggr] .
\end{equation}
This is our expression for the Hamiltonian charge corresponding to the local Lorentz boost and is defined on 
a WIH, through which no matter falls in, and hence represents an isolated black hole horizon. 
\\

Let us now evaluate the expression for $H_\lambda$ in different spacetime dimensions. The number of Lovelock contributions that appear is determined by the upper limit of the summation, $p_{\rm max}=\left[(N-1)/2\right]$.
As the spacetime dimension increases, higher-order Lovelock terms begin to contribute. We illustrate this explicitly up to seven spacetime dimensions.
For $N=3$, we have $p_{\rm max}=1$, and therefore only the Einstein-Hilbert term contributes, so the Hamiltonian reduces to:
\begin{equation}
H_\lambda=\oint_{S_\Delta}
{}^{1}\epsilon\,\alpha_1.
\end{equation}
In particular, one recovers from here that circumference of the horizon is the Hamiltonian charge corresponding to local Lorentz boost in $N=3$ as discussed in section (\ref{2+1}). Thus, in three dimensions, the Hamiltonian reduces to the integral of the one-dimensional volume element, giving the circumference of the horizon.
In four dimensional spacetime , we have $p_{\rm max}=1$, and once again only the first Lovelock term contributes. Consequently,
\begin{equation}
H_\lambda =\oint_{S_\Delta}
{}^{2}\epsilon\,\alpha_1.
\end{equation}
So the Hamiltonian in 4 spacetime dimension is proportional to area of the horizon. This is precisely the familiar Einstein gravity contribution, as expected.
In five dimensions, we have $p_{\rm max}=2$, so both the Einstein-Hilbert and the Gauss-Bonnet terms contribute. The Hamiltonian therefore takes the form:
\begin{equation}
H_\lambda=\oint_{S_\Delta}{}^{3}\epsilon\left[\alpha_1+ \alpha_2\,\delta^{i_1i_2}_{j_1j_2} {\mathcal R_{i_1i_2}}^{j_1j_2}\right].
\end{equation}
Expanding the generalized Kronecker delta we can find that
$\delta^{ab}_{cd}\,{\mathcal R_{ab}}^{cd}=2\mathcal R$,
and hence we get:
\begin{equation}
H_\lambda=\oint_{S_\Delta}{}^{3}\epsilon\left(\alpha_1+2\alpha_2\,\mathcal R\right),
\end{equation}
where $\mathcal R$ is the Ricci scalar of the horizon cross-section. We therefore observe that while only the Einstein-Hilbert contribution survives in three and four dimensions, the Gauss-Bonnet correction starts contributing from five dimensions onwards. Higher-dimensional spacetimes similarly receive contributions progressively from higher-order Lovelock densities. As seen in the transition from three to four dimensions, no new contribution appears in the Hamiltonian when passing from an odd spacetime dimension to the next even dimension; only the induced volume element increases in dimension. Thus, the six-dimensional expression follows directly from the five-dimensional one. This is because the additional Lovelock term is the Euler density, which is topological. So in the six spacetime dimensions the expression for the Hamiltonian is:
\begin{equation}
H_\lambda=\oint_{S_\Delta}{}^{4}\epsilon\left(\alpha_1+2\alpha_2\,\mathcal R\right),
\end{equation}
In $7$ dimensions, $p_{\rm max}=3$, and hence the cubic Lovelock term also contributes. The Hamiltonian therefore takes the form:
\begin{equation}
H_{\lambda}=\oint_{S_\Delta}{}^{5}\epsilon\left[\alpha_1+2\alpha_2\,\mathcal{R}+3\alpha_3\left(\mathcal{R}^2
-4\mathcal{R}_{ab}\mathcal{R}^{ab}
+\mathcal{R}_{abcd}\mathcal{R}^{abcd}\right)\right],
\end{equation}
We note that these are precisely the intrinsic curvature invariants that appear in the Wald entropy formula for Lovelock gravity. Therefore the integrand in equation \eqref{boost charge} is Lanczos-Lovelock scalar on the horizon cross section, which is identical to the expression of Wald entropy associated with the horizon \cite{Iyer:1995kg, Wald:1993nt, Jacobson:1993vj}.

\section{Discussion}\label{Discussions}
In this work, we have shown that the Hamiltonian charge corresponding to the local Lorentz boost is modified geometric area which is identical to the Wald entropy in Lanczos-Lovelock class of theories. However, gauge symmetries are usually generated by zero modes and hence do not give rise to any physical charge, the presence of boundaries renders these gauge symmteries physical which can give rise to non-vanishing charge. In our case this charge is precisely the modified geometric area identical to the Wald entropy. Earlier analysis based on Euclidean action principle established the relation between boost parameter and geometric area \cite{Banados:1993qp,Carlip:1993sa,Carlip:1994gc}.  We believe that our result is realization of the same idea in Lorentzian signature  using completely different methods based on covariant phase space, but now generalized to broader class of gravitational theories. Also the relation between boost energy $K$ and area $A$ of the bifurcation surface for Rindler horizons generated by boost Killing vector field was obtained in \cite{Wall:2010cj} which looks like $A=c-8\pi G K$, for some constant $c$. Remarkably our result in \eqref{boost charge} is mathematically similar to this equation in more general class of gravitational theories where constant $c$ can be inserted as constant of integration.  And consequently it can be argued that the connection between area and boost is not restricted to Rindler horizons and in fact it is an intrinsic character of the horizon independent of the observer. But in our analysis the area arises as Hamiltonian charge due to the local Lorentz boost and not due to the boost Killing field. Moreover the resemblance of near horizon geometry as Rindler spacetime breaks down for extremal black holes. Consequently it is not clear how one can obtain  such a correspondence between boost and area for spacetime containing extremal black holes. Such a problem is present in Wald’s interpretation of black hole entropy as a Noether charge as well, which requires the presence of a bifurcation surface where the Killing field vanishes. Such a bifurcation surface is not available for extremal black holes  and also for realistic black holes formed through gravitational collapse of the matter. However the phase space of isolated horizons can be extended to include extremal black hole solutions \cite{Chatterjee:2008if}. And since we have used isolated horizon to describe black hole boundary,  our result continues  to hold for extremal black holes as well. One can also view the local Lorentz symmetry on the horizon to be equivalent to diffeomorphism generated by boost Killing field of a local Rindler observer. The argument goes as follows: Let  $\xi^\mu $ be the representative of isometries on spacetime , that implies $\lie_\xi g_{ab}=0$. But this does not imply that the $\lie_\xi e^I_a=0$ as they can have rotation in the internal space while being dragged by $\xi^\mu$. In fact, one can show if $\lie_\xi e^I_a={\epsilon_{(\xi)}}^I_J e^J_a$, then from here it follows that $\lie_{\xi}g_{ab}=0$ (This can be presented consistently if one demands that $e^I_a$ are dragged along $\xi^\mu$ by modified derivative called Lorentz-Lie derivative $\mathcal{K}_{\xi}$ \cite{Jacobson:2015uqa}). This relation suggests that the diffeomorphisms generated by boost Killing field are realized as local Lorentz boost on the tetrad fields. From this point of view one can relate the Hamiltonian charge corresponding to local Lorentz boost on the horizon to boost energy corresponding to Killing field associated with local Rindler observer. Taken together, in this work we suggest that the Wald entropy can be identified as a Hamiltonian charge corresponding to the boost, but to place the claim on more firm footing, generalization of this to arbitrary theories of gravity derived from diffeomorphism invariant Lagrangian is necessary.  For instance, an interesting direction for future work would be to investigate whether the present analysis can be extended to Horndeski gravity \cite{Horndeski:1974wa}. In this work, we have considered black holes in equilibrium modeled by isolated horizons, but further generalization of this work to dynamical settings \cite{Iyer:1994ys,Dong:2013qoa,Wall:2015raa,Bhattacharyya:2021jhr,Hollands:2024vbe,Bhattacharyya:2026aiu}, where the ambiguities at various levels in the construction of the Noether charge can affect the definition of entropy \cite{Jacobson:1993vj}, will be investigated and presented in future work.\\

Finally, note that since the Wald entropy is identified from variational equation, it is ambiguous up to addition of terms that do not vary over phase space. In fact the variations that one considers in this technique cannot take one solution to another in which the black hole topology is different. This therefore may or may not give rise to appearance of topological invariants in the expression for Wald entropy when topological terms are added to the action. Consequently, the Wald entropy may differ from the entropy calculated using other methods \cite{Krishnan:2009tj}. It is to be noted that our result may also differ from Wald entropy by the appearance of such topological terms.

\section{Acknowledgements} S.D. acknowledges the financial support provided by DST vide Grant No.DST/INSPIRE Fellowship/2020/IF200537. S.D. also acknowledges useful discussions with Sk Jahanur Hoque and Swastik Bhattacharya during the final stages of this work. Both Ayan Chatterjee and Amit Ghosh acknowledge the support of  DAE-BRNS project $58/14/25/2019$-BRNS.  A.C. and S.D. acknowledge fruitful discussions with K.P. Yogendran, V. Pravda, M. Ortaggio, Nemani Suryanarayana, Kavita Kumari, and Eshita Biswas.  
\section{Appendix}
\subsection{Riemann and Weyl 
tensor}\label{Appendix A}
We can expand Riemann tensor in the chosen basis vectors as follows:
\begin{eqnarray} \label{Riemann expansion}
R_{abcd} &=& 4 R_{0i0j} \, n_{\{a} m^{(i)}_{b} n_{c} m^{(j)}_{d\}} 
+ 8 R_{010i} \, n_{\{a} \ell_{b} n_{c} m^{(i)}_{d\}} 
+ 4 R_{0ijk} \, n_{\{a} m^{(i)}_{b} m^{(j)}_{c} m^{(k)}_{d\}} \nonumber \\
&& + 4 R_{0101} \, n_{\{a} \ell_{b} n_{c} \ell_{d\}} 
+ 4 R_{01ij} \, n_{\{a} \ell_{b} m^{(i)}_{c} m^{(j)}_{d\}} 
+ 8 R_{0i1j} \, n_{\{a} m^{(i)}_{b} \ell_{c} m^{(j)}_{d\}} \nonumber \\
&& + R_{ijkl} \, m^{(i)}_{\{a} m^{(j)}_{b} m^{(k)}_{c} m^{(l)}_{d\}} 
+ 8 R_{101i} \, \ell_{\{a} n_{b} \ell_{c} m^{(i)}_{d\}} \nonumber + 4 R_{1ijk} \, \ell_{\{a} m^{(i)}_{b} m^{(j)}_{c} m^{(k)}_{d\}} 
\\
&& + 4 R_{1i1j} \, \ell_{\{a} m^{(i)}_{b} \ell_{c} m^{(j)}_{d\}} .
\end{eqnarray}
where the bracket $\{~\}$ is defined as :
\begin{equation}
    P_{\{a}Q_b R_c S_{d\}}:=\frac{1}{2}(P_{[a}Q_{b]} R_{[c} S_{d]}+P_{[c}Q_{d]} R_{[a} S_{b]})
\end{equation}
Similarly we can expand the Ricci tensor as follows:
\begin{eqnarray}\label{Ricci expansion}
    R_{ab}&=& R_{0i0i}n_a n_b +2(R_{010i}+R_{0k ik})n_{(a}m_{b)}^{(i)}+2(R_{0101}+R_{0i1i})n_{(a}\ell_{b)}+2(R_{kikj}- R_{0i1j})m_{(a}^{(i)}m_{b)}^{(j)}\\ \nonumber
    && 2(R_{101i}+R_{1kik})\ell_a m_b^{(i)} + R_{1i1i} \ell_a \ell_b
\end{eqnarray}

so that,
\begin{equation}
R_{\underleftarrow{ab}}\stackrel{\Delta}{=}2(R_{kikj} -R_{0i1j} )m_{(a}^{(i)} m_{b)}^{(j)}
\end{equation}

In its most general form, the Weyl tensor admits the following decomposition with respect to the chosen basis:
\begin{eqnarray} \label{Weyl expansion}
C_{abcd} &=& 4 C_{0i0j} \, n_{\{a} m^{(i)}_{b} n_{c} m^{(j)}_{d\}} 
+ 8 C_{010i} \, n_{\{a} \ell_{b} n_{c} m^{(i)}_{d\}} 
+ 4 C_{0ijk} \, n_{\{a} m^{(i)}_{b} m^{(j)}_{c} m^{(k)}_{d\}} \nonumber \\
&& + 4 C_{0101} \, n_{\{a} \ell_{b} n_{c} \ell_{d\}} 
+ 4 C_{01ij} \, n_{\{a} \ell_{b} m^{(i)}_{c} m^{(j)}_{d\}} 
+ 8 C_{0i1j} \, n_{\{a} m^{(i)}_{b} \ell_{c} m^{(j)}_{d\}} \nonumber \\
&& + C_{ijkl} \, m^{(i)}_{\{a} m^{(j)}_{b} m^{(k)}_{c} m^{(l)}_{d\}} 
+ 8 C_{101i} \, \ell_{\{a} n_{b} \ell_{c} m^{(i)}_{d\}} \nonumber + 4 C_{1ijk} \, \ell_{\{a} m^{(i)}_{b} m^{(j)}_{c} m^{(k)}_{d\}} 
\\
&& + 4 C_{1i1j} \, \ell_{\{a} m^{(i)}_{b} \ell_{c} m^{(j)}_{d\}} .
\end{eqnarray}

\subsection{Bulk contribution to the symplectic structure}\label{A2}
To prove \eqref{bulk symplectic structure}, let us begin with \eqref{Sigma 1} and \eqref{Sigma p}. We start by writing \eqref{Sigma p} in slightly different form as follows:
\begin{equation}
    \Sigma^p_{IJ}=\Sigma_{IJ a_3 a_4\cdots a_{2p-1} a_{2p}} \wedge F^{a_3 a_4}\wedge \cdots \wedge F^{a_{2p-1} a_{2p}}
\end{equation}
where,
\begin{equation}
    \Sigma_{IJ a_3 a_4\cdots a_{2p-1} a_{2p}}=\frac{1}{(N-2p)!}\, \epsilon_{IJ a_3 a_4 \cdots a_{2p} a_{2p+1}\cdots a_N} e^{a_{2p+1}}\wedge\cdots\wedge e^{a_N}
\end{equation}
Let us first calculate the term $\delta_{\lambda} \Tilde{\Sigma}_{IJ} \wedge \delta A^{IJ}$. 
\begin{eqnarray*}
    \delta_\lambda \Tilde{\Sigma}_{IJ}=\sum_{p=1}^{\left[\dfrac{N-1}{2}\right]} p\alpha_p\, \delta_\lambda(\Sigma_{IJ} ^p) 
\end{eqnarray*}
\begin{eqnarray} 
    \delta_\lambda(\Sigma_{IJ} ^p) =\delta_\lambda (\Sigma_{IJ a_3 a_4 \cdots a_{2p-1}a_{2p}})\wedge F^{a_3 a_4}\wedge\cdots\wedge F^{a_{2p-1} a_{2p}}+ \Sigma_{IJ a_3 a_4 \cdots a_{2p-1}a_{2p}}\wedge \delta_\lambda(F^{a_3 a_4}\wedge\cdots\wedge F^{a_{2p-1} a_{2p}})
\end{eqnarray}
\begin{eqnarray}\nonumber
    \delta_\lambda (\Sigma_{IJ a_3 a_4 \cdots a_{2p-1}a_{2p}})&=&\frac{1}{(N-2p)!}\,\delta_\lambda(\epsilon_{IJ a_3 a_4\cdots a_{2p} a_{2p+1}\cdots a_N \, e^{a_{2p+1}}\wedge\cdots e^{a_N}})\\ \nonumber
    &=& \frac{N-2p}{(N-2p)!}\epsilon_{IJ a_3 a_4\cdots a_{2p} a_{2p+1}\cdots a_N} \, e^{a_{2p+1}}\wedge \cdots \delta_\lambda e^{a_N}\\ \label{A}
    &=& \frac{1}{(N-2p-1)!}\epsilon_{IJ a_3 a_4\cdots a_{2p} a_{2p+1}\cdots a_N} \,{\epsilon^{a_N}}_N \, e^{a_{2p+1}}\wedge \cdots \wedge e^{a_{N-1}}\wedge e^N
\end{eqnarray} 
We can use  $\epsilon_{\nu_1 \nu_2\cdots \nu_p \alpha_1\cdots \alpha_{N-p}}\epsilon^{\nu_1 \nu_2\cdots \nu_p \beta_1\cdots \beta_{N-p}}=-p! (N-p)! \delta^{[\beta_1}_{\alpha_1}\cdots \delta^{\beta_{N-p}]}_{\alpha_{N-p}}$ to show that:
\begin{equation}\label{B}
    e^{a_{2p+1}}\wedge e^{a_{2p+2}}\wedge \cdots \wedge e^{a_{N-1}}\wedge e^N=-\frac{1}{(2p)!} \epsilon ^{KL c_3 c_4\cdots c_{2p} c_{2p+1}\cdots a_{N-1} N}\,\Sigma_{KL c_3\cdots c_{2p}}
\end{equation}
Using \eqref{B} in \eqref{A} it is easy to show that:
\begin{eqnarray}\label{C}
    \delta_{\lambda}(\Sigma_{IJ a_3 a_4 \cdots a_{2p-1} a_{2p}})=\frac{(2p+1)!}{2p!} \delta^K_{[I} \delta^L_{J} \delta^{c_3}_{a_3}\cdots \delta^{c_{2p}}_{a_{2p}} \delta^N_{a_{N}]}\Sigma_{KL c_3\cdots c_{2p}}{\epsilon^{a_N}}_N
\end{eqnarray}
We write \eqref{C} as follows:
\begin{eqnarray} \nonumber
    \delta_{\lambda}(\Sigma_{IJ a_3 a_4 \cdots a_{2p-1} a_{2p}})=&\frac{1}{2p!}&\{2p! \,\delta^K_{[I} \delta^L_{J} \delta^{c_3}_{a_3}\cdots \delta^{c_{2p-1}}_{a_{2p-1}}\delta^{c_{2p}}_{a_{2p}]} \delta^N_{a_{N}}-2p!\delta^K_{[a_{N}} \delta^L_{J} \delta^{c_3}_{a_3}\cdots \delta^{c_{2p-1}}_{a_{2p-1}}\delta^{c_{2p}}_{a_{2p}]} \delta^N_{I}\\ \nonumber
     -&2p!&\delta^K_{[I} \delta^L_{a_N} \delta^{c_3}_{a_3}\cdots \delta^{c_{2p-1}}_{a_{2p-1}}\delta^{c_{2p}}_{a_{2p}]} \delta^N_{J}-2p!\delta^K_{[I} \delta^L_{J} \delta^{c_3}_{a_N}\cdots \delta^{c_{2p-1}}_{a_{2p-1}}\delta^{c_{2p}}_{a_{2p}]} \delta^N_{a_3}-\cdots \cdots\\ 
     -&2p!&\delta^K_{[I} \delta^L_{J} \delta^{c_3}_{a_3}\cdots \delta^{c_{2p-1}}_{a_{N}}\delta^{c_{2p}}_{a_{2p}]} \delta^N_{a_{2p-1}}
     -2p!\delta^K_{[I} \delta^L_{J} \delta^{c_3}_{a_3}\cdots \delta^{c_{2p-1}}_{a_{2p-1}}\delta^{c_{2p}}_{a_{N}]} \delta^N_{a_{2p}}\}\, \Sigma_{KL c_3\cdots c_{2p}} {\epsilon^{a_N}}_N
\end{eqnarray}
\begin{eqnarray}\nonumber
    \delta_{\lambda}(\Sigma_{IJ a_3 a_4 \cdots a_{2p-1} a_{2p}})&=&-\{\Sigma_{a_N Ja_3 a_4\cdots a_{2p-1} a_{2p}}\, {\epsilon^{a_N}}_I+ \Sigma_{I a_N a_3 a_4\cdots a_{2p-1} a_{2p}}\, {\epsilon^{a_N}}_J\\ \nonumber
    &+&\Sigma_{I Ja_N a_4\cdots a_{2p-1} a_{2p}}\, {\epsilon^{a_N}}_{a_{3}} +\Sigma_{I Ja_3 a_N\cdots a_{2p-1} a_{2p}}\, {\epsilon^{a_N}}_{a_4}+\cdots\\
    &\cdots&+ \Sigma_{I Ja_3 a_4\cdots a_{N} a_{2p}}\, {\epsilon^{a_N}}_{a_{2p-1}} +\Sigma_{I Ja_3 a_4\cdots a_{2p-1} a_{N}}\, {\epsilon^{a_N}}_{a_{2p}}\}
\end{eqnarray}
Invoking the antisymmetry in $I,J$, which will arise once the expression is contracted with $A^{IJ}$, and the antisymmetry among the $a_i$'s, which will appear upon multiplication with $ F^{a_3 a_4} \wedge \cdots \wedge F^{a_{2p-1} a_{2p}},$ it is easy to show that:
\begin{eqnarray}\label{D}
    \delta_\lambda(\Sigma_{IJa_3\cdots a_{2p}}) \wedge F^{a_3 a_4}\wedge\cdots \wedge F^{a_{2p-1} a_{2p}}\wedge \delta A^{IJ}=-&2&\Sigma_{a_N Ja_3 a_4\cdots a_{2p-1} a_{2p}}\, {\epsilon^{a_N}}_I\wedge F^{a_3 a_4}\wedge\cdots \wedge F^{a_{2p-1} a_{2p}}\wedge \delta A^{IJ}\\ \nonumber
    -&2&(p-1)\Sigma_{I Ja_3 a_4\cdots a_{2p-1} a_{N}}\, {\epsilon^{a_N}}_{a_{2p}}\wedge F^{a_3 a_4}\wedge\cdots \wedge F^{a_{2p-1} a_{2p}}\wedge \delta A^{IJ}
\end{eqnarray}
Next we can also show that:
\begin{eqnarray}\label{E}
    \Sigma_{IJa_3 a_4\cdots a_{2p}}\wedge \delta_{\lambda}(F^{a_3 a_4}\wedge\cdots \wedge F^{a_{2p-1}a_{2p}})\wedge \delta A^{IJ}=2(p-1)\Sigma_{IJa_3\cdots a_{2p}}\wedge F^{a_3 a_4}\wedge \cdots F^{a_{2p-1}M} {\epsilon^{2p}}_M \wedge \delta A^{IJ}
\end{eqnarray}
From \eqref{D} and \eqref{E} we get:
\begin{eqnarray}
    \delta_{\lambda} \Sigma^p_{IJ}\,\wedge\, \delta A^{IJ}=-2\Sigma_{a_N Ja_3 a_4\cdots a_{2p-1} a_{2p}}\, {\epsilon^{a_N}}_I\wedge F^{a_3 a_4}\wedge\cdots \wedge F^{a_{2p-1} a_{2p}}\wedge \delta A^{IJ}
\end{eqnarray}
\begin{eqnarray} \nonumber
    \delta_\lambda\Tilde{\Sigma}_{IJ}\wedge \delta A^{IJ}&=&-2\sum_{p=1}^{\left[\frac{N-1}{2}\right]} p\alpha_p \Sigma_{a_N Ja_3 a_4\cdots a_{2p-1} a_{2p}}\, {\epsilon^{a_N}}_I\wedge F^{a_3 a_4}\wedge\cdots \wedge F^{a_{2p-1} a_{2p}}\wedge \delta A^{IJ}\\ \nonumber
    &=& -2\sum_{p=1}^{\left[\frac{N-1}{2}\right]} p\alpha_p \Sigma ^p_{a_NJ} {\epsilon^{a_N}}_I \wedge \delta A^{IJ}\\ \label{F} 
    &=& -2 \Tilde{\Sigma}_{a_N J} \,{\epsilon^{a_N}}_I \wedge \delta A^{IJ}
\end{eqnarray}
The calculation of $\delta \Tilde{\Sigma}_{IJ}\wedge \delta_\lambda A^{IJ}$ is very simple and goes as follows:
\begin{eqnarray}\nonumber
   \delta \Tilde{\Sigma}_{IJ}\wedge \delta_\lambda A^{IJ}&=& \delta \Tilde{\Sigma}_{IJ} \wedge({A^I}_L \epsilon^{JL} + \epsilon^{IL} {A_{L}}^J-d\epsilon^{IJ})\\ \nonumber
   &=& 2\delta \Tilde{\Sigma}_{IJ} \wedge {A^I}_L \epsilon^{JL}-\delta\Tilde{\Sigma}_{IJ}\wedge d\epsilon  ^{IJ}\\ \nonumber
   &=& 2\delta \Tilde{\Sigma}_{IJ} \wedge {A^I}_L \epsilon^{JL} -(-1)^N d(\epsilon^{IJ} \delta \Tilde{\Sigma}_{IJ})+(-1)^N d(\delta \Tilde{\Sigma}_{IJ}) \epsilon^{IJ}\\ \nonumber
   &=&  2\delta \Tilde{\Sigma}_{IJ} \wedge {A^I}_L \epsilon^{JL} -(-1)^N d(\epsilon^{IJ} \delta \Tilde{\Sigma}_{IJ})+2\delta \Tilde{\Sigma}_{IK}\wedge {A^K}_J \epsilon^{IJ}+2 \Tilde{\Sigma}_{IK}\wedge \delta{A^K}_J \epsilon^{IJ}\\ \label {G}
   &=&2 \Tilde{\Sigma}_{IK}\wedge \delta{A^K}_J \epsilon^{IJ}-(-1)^N d(\epsilon^{IJ} \delta \Tilde{\Sigma}_{IJ})
\end{eqnarray}
where in fourth line we have used equation of motion \eqref{connection eom}.
From \eqref{F} and \eqref{G} it follows that:
\begin{equation}
    (-1)^N\, (\delta \Tilde{\Sigma}_{IJ} \wedge \delta_\lambda A^{IJ} - \delta_\lambda\Tilde{\Sigma}_{IJ} \wedge \delta A^{IJ})=-d(\epsilon^{IJ} \delta \Tilde{\Sigma} _{IJ})
\end{equation}
\bibliographystyle{apsrev4-2}
\bibliography{references}

@article{Bardeen:1973gs,
    author = "Bardeen, James M. and Carter, B. and Hawking, S. W.",
    title = "{The Four laws of black hole mechanics}",
    doi = "10.1007/BF01645742",
    journal = "Commun. Math. Phys.",
    volume = "31",
    pages = "161--170",
    year = "1973"
}

@article{Gibbons:1977mu,
    author = "Gibbons, G. W. and Hawking, S. W.",
    title = "{Cosmological Event Horizons, Thermodynamics, and Particle Creation}",
    doi = "10.1103/PhysRevD.15.2738",
    journal = "Phys. Rev. D",
    volume = "15",
    pages = "2738--2751",
    year = "1977"
}

@article{Hawking:1974rv,
    author = "Hawking, S. W.",
    title = "{Black hole explosions}",
    doi = "10.1038/248030a0",
    journal = "Nature",
    volume = "248",
    pages = "30--31",
    year = "1974"
}

@article{Hawking:1975vcx,
    author = "Hawking, S. W.",
    editor = "Gibbons, G. W. and Hawking, S. W.",
    title = "{Particle Creation by Black Holes}",
    doi = "10.1007/BF02345020",
    journal = "Commun. Math. Phys.",
    volume = "43",
    pages = "199--220",
    year = "1975",
    note = "[Erratum: Commun.Math.Phys. 46, 206 (1976)]"
}

@article{Bekenstein:1972tm,
    author = "Bekenstein, J. D.",
    title = "{Black holes and the second law}",
    doi = "10.1007/BF02757029",
    journal = "Lett. Nuovo Cim.",
    volume = "4",
    pages = "737--740",
    year = "1972"
}

@article{Bekenstein:1973ur,
    author = "Bekenstein, Jacob D.",
    title = "{Black holes and entropy}",
    doi = "10.1103/PhysRevD.7.2333",
    journal = "Phys. Rev. D",
    volume = "7",
    pages = "2333--2346",
    year = "1973"
}

@article{Lee:1990nz,
    author = "Lee, J. and Wald, Robert M.",
    title = "{Local symmetries and constraints}",
    doi = "10.1063/1.528801",
    journal = "J. Math. Phys.",
    volume = "31",
    pages = "725--743",
    year = "1990"
}

@article{Wald:1993nt,
    author = "Wald, Robert M.",
    title = "{Black hole entropy is the Noether charge}",
    eprint = "gr-qc/9307038",
    archivePrefix = "arXiv",
    reportNumber = "EFI-93-42",
    doi = "10.1103/PhysRevD.48.R3427",
    journal = "Phys. Rev. D",
    volume = "48",
    number = "8",
    pages = "R3427--R3431",
    year = "1993"
}

@article{Iyer:1994ys,
    author = "Iyer, Vivek and Wald, Robert M.",
    title = "{Some properties of Noether charge and a proposal for dynamical black hole entropy}",
    eprint = "gr-qc/9403028",
    archivePrefix = "arXiv",
    doi = "10.1103/PhysRevD.50.846",
    journal = "Phys. Rev. D",
    volume = "50",
    pages = "846--864",
    year = "1994"
}

@article{Carlip:1993sa,
    author = "Carlip, Steven and Teitelboim, Claudio",
    title = "{The Off-shell black hole}",
    eprint = "gr-qc/9312002",
    archivePrefix = "arXiv",
    reportNumber = "IASSNS-HEP-93-84, UCD-93-34",
    doi = "10.1088/0264-9381/12/7/011",
    journal = "Class. Quant. Grav.",
    volume = "12",
    pages = "1699--1704",
    year = "1995"
}

@article{Strominger:1996sh,
    author = "Strominger, Andrew and Vafa, Cumrun",
    title = "{Microscopic origin of the Bekenstein-Hawking entropy}",
    eprint = "hep-th/9601029",
    archivePrefix = "arXiv",
    reportNumber = "HUTP-96-A002, RU-96-01",
    doi = "10.1016/0370-2693(96)00345-0",
    journal = "Phys. Lett. B",
    volume = "379",
    pages = "99--104",
    year = "1996"
}

@inproceedings{Peet:2000hn,
    author = "Peet, Amanda W.",
    title = "{TASI lectures on black holes in string theory}",
    booktitle = "{Theoretical Advanced Study Institute in Elementary Particle Physics (TASI 99): Strings, Branes, and Gravity}",
    eprint = "hep-th/0008241",
    archivePrefix = "arXiv",
    reportNumber = "UTPT-00-10, NSF-ITP-99-145",
    doi = "10.1142/9789812799630_0003",
    pages = "353--433",
    month = "8",
    year = "2000"
}

@article{Rovelli:1996dv,
    author = "Rovelli, Carlo",
    title = "{Black hole entropy from loop quantum gravity}",
    eprint = "gr-qc/9603063",
    archivePrefix = "arXiv",
    doi = "10.1103/PhysRevLett.77.3288",
    journal = "Phys. Rev. Lett.",
    volume = "77",
    pages = "3288--3291",
    year = "1996"
}

@article{Ashtekar:1997yu,
    author = "Ashtekar, A. and Baez, J. and Corichi, A. and Krasnov, Kirill",
    title = "{Quantum geometry and black hole entropy}",
    eprint = "gr-qc/9710007",
    archivePrefix = "arXiv",
    reportNumber = "CGPG-97-9-3",
    doi = "10.1103/PhysRevLett.80.904",
    journal = "Phys. Rev. Lett.",
    volume = "80",
    pages = "904--907",
    year = "1998"
}

@article{Kaul:1998xv,
    author = "Kaul, Romesh K. and Majumdar, Parthasarathi",
    title = "{Quantum black hole entropy}",
    eprint = "gr-qc/9801080",
    archivePrefix = "arXiv",
    reportNumber = "IMSC-97-11-39",
    doi = "10.1016/S0370-2693(98)01030-2",
    journal = "Phys. Lett. B",
    volume = "439",
    pages = "267--270",
    year = "1998"
}

@article{Domagala:2004jt,
    author = "Domagala, Marcin and Lewandowski, Jerzy",
    title = "{Black hole entropy from quantum geometry}",
    eprint = "gr-qc/0407051",
    archivePrefix = "arXiv",
    doi = "10.1088/0264-9381/21/22/014",
    journal = "Class. Quant. Grav.",
    volume = "21",
    pages = "5233--5244",
    year = "2004"
}

@article{Meissner:2004ju,
    author = "Meissner, Krzysztof A.",
    title = "{Black hole entropy in loop quantum gravity}",
    eprint = "gr-qc/0407052",
    archivePrefix = "arXiv",
    doi = "10.1088/0264-9381/21/22/015",
    journal = "Class. Quant. Grav.",
    volume = "21",
    pages = "5245--5252",
    year = "2004"
}

@article{Ghosh:2004wq,
    author = "Ghosh, A. and Mitra, P.",
    title = "{An Improved lower bound on black hole entropy in the quantum geometry approach}",
    eprint = "gr-qc/0411035",
    archivePrefix = "arXiv",
    doi = "10.1016/j.physletb.2005.05.003",
    journal = "Phys. Lett. B",
    volume = "616",
    pages = "114--117",
    year = "2005"
}

@article{Engle:2009vc,
    author = "Engle, Jonathan and Perez, Alejandro and Noui, Karim",
    title = "{Black hole entropy and SU(2) Chern-Simons theory}",
    eprint = "0905.3168",
    archivePrefix = "arXiv",
    primaryClass = "gr-qc",
    doi = "10.1103/PhysRevLett.105.031302",
    journal = "Phys. Rev. Lett.",
    volume = "105",
    pages = "031302",
    year = "2010"
}

@article{Kaul:2000kf,
    author = "Kaul, Romesh K. and Majumdar, Parthasarathi",
    title = "{Logarithmic correction to the Bekenstein-Hawking entropy}",
    eprint = "gr-qc/0002040",
    archivePrefix = "arXiv",
    doi = "10.1103/PhysRevLett.84.5255",
    journal = "Phys. Rev. Lett.",
    volume = "84",
    pages = "5255--5257",
    year = "2000"
}

@article{Chatterjee:2020iuf,
    author = "Chatterjee, Ayan and Ghosh, Amit",
    title = "{Exponential Corrections to Black Hole Entropy}",
    eprint = "2007.15401",
    archivePrefix = "arXiv",
    primaryClass = "gr-qc",
    doi = "10.1103/PhysRevLett.125.041302",
    journal = "Phys. Rev. Lett.",
    volume = "125",
    number = "4",
    pages = "041302",
    year = "2020"
}

@article{Garg:2026toh,
    author = "Garg, Akriti and Devdutt, Sahil and Chatterjee, Ayan",
    title = "{Nonminimally coupled scalar field, area quantization, and black hole entropy}",
    eprint = "2603.25292",
    archivePrefix = "arXiv",
    primaryClass = "gr-qc",
    doi = "10.1103/wb82-sq14",
    journal = "Phys. Rev. D",
    volume = "114",
    number = "2",
    pages = "024006",
    year = "2026"
}

@article{Dabholkar:2004yr,
    author = "Dabholkar, Atish",
    title = "{Exact counting of black hole microstates}",
    eprint = "hep-th/0409148",
    archivePrefix = "arXiv",
    reportNumber = "SLAC-PUB-10717, SU-ITP-04-36, TIFR-TH-04-23",
    doi = "10.1103/PhysRevLett.94.241301",
    journal = "Phys. Rev. Lett.",
    volume = "94",
    pages = "241301",
    year = "2005"
}

@article{Dabholkar:2005dt,
    author = "Dabholkar, Atish and Denef, Frederik and Moore, Gregory W. and Pioline, Boris",
    title = "{Precision counting of small black holes}",
    eprint = "hep-th/0507014",
    archivePrefix = "arXiv",
    reportNumber = "LPTHE-05-14, LPTENS-05-21, TIFR-TH-05-27",
    doi = "10.1088/1126-6708/2005/10/096",
    journal = "JHEP",
    volume = "10",
    pages = "096",
    year = "2005"
}

@article{Dijkgraaf:1996it,
    author = "Dijkgraaf, Robbert and Verlinde, Erik P. and Verlinde, Herman L.",
    title = "{Counting dyons in N=4 string theory}",
    eprint = "hep-th/9607026",
    archivePrefix = "arXiv",
    reportNumber = "CERN-TH-96-170",
    doi = "10.1016/S0550-3213(96)00640-2",
    journal = "Nucl. Phys. B",
    volume = "484",
    pages = "543--561",
    year = "1997"
}

@article{Mandal:2010cj,
    author = "Mandal, Ipsita and Sen, Ashoke",
    editor = "Baulieu, Laurent and de Boer, Jan and Douglas, Michael and Rabinovici, Eliezer and Vanhove, Pierre and Windey, Paul",
    title = "{Black Hole Microstate Counting and its Macroscopic Counterpart}",
    eprint = "1008.3801",
    archivePrefix = "arXiv",
    primaryClass = "hep-th",
    doi = "10.1088/0264-9381/27/21/214003",
    journal = "Class. Quant. Grav.",
    volume = "27",
    pages = "214003",
    year = "2010"
}

@article{Gibbons:1976ue,
    author = "Gibbons, G. W. and Hawking, S. W.",
    title = "{Action Integrals and Partition Functions in Quantum Gravity}",
    reportNumber = "PRINT-76-0995 (CAMBRIDGE)",
    doi = "10.1103/PhysRevD.15.2752",
    journal = "Phys. Rev. D",
    volume = "15",
    pages = "2752--2756",
    year = "1977"
}

@article{Iyer:1995kg,
    author = "Iyer, Vivek and Wald, Robert M.",
    title = "{A Comparison of Noether charge and Euclidean methods for computing the entropy of stationary black holes}",
    eprint = "gr-qc/9503052",
    archivePrefix = "arXiv",
    doi = "10.1103/PhysRevD.52.4430",
    journal = "Phys. Rev. D",
    volume = "52",
    pages = "4430--4439",
    year = "1995"
}

@article{Bombelli:1986rw,
    author = "Bombelli, Luca and Koul, Rabinder K. and Lee, Joohan and Sorkin, Rafael D.",
    title = "{A Quantum Source of Entropy for Black Holes}",
    reportNumber = "PRINT-86-0371 (SYRACUSE)",
    doi = "10.1103/PhysRevD.34.373",
    journal = "Phys. Rev. D",
    volume = "34",
    pages = "373--383",
    year = "1986"
}

@article{Callan:1994py,
    author = "Callan, Jr., Curtis G. and Wilczek, Frank",
    title = "{On geometric entropy}",
    eprint = "hep-th/9401072",
    archivePrefix = "arXiv",
    reportNumber = "IASSNS-HEP-93-87",
    doi = "10.1016/0370-2693(94)91007-3",
    journal = "Phys. Lett. B",
    volume = "333",
    pages = "55--61",
    year = "1994"
}

@article{Solodukhin:2011gn,
    author = "Solodukhin, Sergey N.",
    title = "{Entanglement entropy of black holes}",
    eprint = "1104.3712",
    archivePrefix = "arXiv",
    primaryClass = "hep-th",
    doi = "10.12942/lrr-2011-8",
    journal = "Living Rev. Rel.",
    volume = "14",
    pages = "8",
    year = "2011"
}

@article{Holzhey:1994we,
    author = "Holzhey, Christoph and Larsen, Finn and Wilczek, Frank",
    title = "{Geometric and renormalized entropy in conformal field theory}",
    eprint = "hep-th/9403108",
    archivePrefix = "arXiv",
    reportNumber = "PUPT-1454, IASSNS-HEP-93-88",
    doi = "10.1016/0550-3213(94)90402-2",
    journal = "Nucl. Phys. B",
    volume = "424",
    pages = "443--467",
    year = "1994"
}

@article{Srednicki:1993im,
    author = "Srednicki, Mark",
    title = "{Entropy and area}",
    eprint = "hep-th/9303048",
    archivePrefix = "arXiv",
    reportNumber = "LBL-33754, CFPA-93-02",
    doi = "10.1103/PhysRevLett.71.666",
    journal = "Phys. Rev. Lett.",
    volume = "71",
    pages = "666--669",
    year = "1993"
}

@article{Frolov:1993ym,
    author = "Frolov, Valeri P. and Novikov, Igor",
    title = "{Dynamical origin of the entropy of a black hole}",
    eprint = "gr-qc/9309001",
    archivePrefix = "arXiv",
    reportNumber = "NORDITA-93-45A",
    doi = "10.1103/PhysRevD.48.4545",
    journal = "Phys. Rev. D",
    volume = "48",
    pages = "4545--4551",
    year = "1993"
}

@article{tHooft:1984kcu,
    author = "'t Hooft, Gerard",
    title = "{On the Quantum Structure of a Black Hole}",
    reportNumber = "Print-84-0924 (UTRECHT)",
    doi = "10.1016/0550-3213(85)90418-3",
    journal = "Nucl. Phys. B",
    volume = "256",
    pages = "727--745",
    year = "1985"
}

@article{Carlip:1999cy,
    author = "Carlip, Steven",
    title = "{Entropy from conformal field theory at Killing horizons}",
    eprint = "gr-qc/9906126",
    archivePrefix = "arXiv",
    reportNumber = "UCD-99-12, NSF-ITP-99-069",
    doi = "10.1088/0264-9381/16/10/322",
    journal = "Class. Quant. Grav.",
    volume = "16",
    pages = "3327--3348",
    year = "1999"
}

@article{Brown:1986nw,
    author = "Brown, J. David and Henneaux, M.",
    title = "{Central Charges in the Canonical Realization of Asymptotic Symmetries: An Example from Three-Dimensional Gravity}",
    doi = "10.1007/BF01211590",
    journal = "Commun. Math. Phys.",
    volume = "104",
    pages = "207--226",
    year = "1986"
}

@article{Strominger:1997eq,
    author = "Strominger, Andrew",
    title = "{Black hole entropy from near horizon microstates}",
    eprint = "hep-th/9712251",
    archivePrefix = "arXiv",
    reportNumber = "HUTP-97-A106",
    doi = "10.1088/1126-6708/1998/02/009",
    journal = "JHEP",
    volume = "02",
    pages = "009",
    year = "1998"
}

@article{Carlip:1994gy,
    author = "Carlip, Steven",
    title = "{The Statistical mechanics of the (2+1)-dimensional black hole}",
    eprint = "gr-qc/9409052",
    archivePrefix = "arXiv",
    reportNumber = "UCD-94-32, NI-94011",
    doi = "10.1103/PhysRevD.51.632",
    journal = "Phys. Rev. D",
    volume = "51",
    pages = "632--637",
    year = "1995"
}

@article{Cardy:1986ie,
    author = "Cardy, John L.",
    title = "{Operator Content of Two-Dimensional Conformally Invariant Theories}",
    doi = "10.1016/0550-3213(86)90552-3",
    journal = "Nucl. Phys. B",
    volume = "270",
    pages = "186--204",
    year = "1986"
}

@article{Blote:1986qm,
    author = "Bl{\"o}te, H. W. J. and Cardy, J. L. and Nightingale, M. P.",
    title = "{Conformal Invariance, the Central Charge and Universal Finite Size Amplitudes at Criticality}",
    journal = "Phys. Rev. Lett.",
    volume = "56",
    pages = "742--745",
    year = "1986",
    doi = "10.1103/PhysRevLett.56.742"
}

@article{Dreyer:2001py,
    author = "Dreyer, Olaf and Ghosh, Amit and Wisniewski, Jacek",
    title = "{Black hole entropy calculations based on symmetries}",
    eprint = "hep-th/0101117",
    archivePrefix = "arXiv",
    reportNumber = "CGPG-01-1-1",
    doi = "10.1088/0264-9381/18/10/308",
    journal = "Class. Quant. Grav.",
    volume = "18",
    pages = "1929--1938",
    year = "2001"
}

@article{Wall:2015raa,
    author = "Wall, Aron C.",
    title = "{A Second Law for Higher Curvature Gravity}",
    eprint = "1504.08040",
    archivePrefix = "arXiv",
    primaryClass = "gr-qc",
    doi = "10.1142/S0218271815440149",
    journal = "Int. J. Mod. Phys. D",
    volume = "24",
    number = "12",
    pages = "1544014",
    year = "2015"
}

@article{Chatterjee:2011wj,
    author = "Chatterjee, Ayan and Sarkar, Sudipta",
    title = "{Physical process first law and increase of horizon entropy for black holes in Einstein-Gauss-Bonnet gravity}",
    eprint = "1111.3021",
    archivePrefix = "arXiv",
    primaryClass = "gr-qc",
    doi = "10.1103/PhysRevLett.108.091301",
    journal = "Phys. Rev. Lett.",
    volume = "108",
    pages = "091301",
    year = "2012"
}

@article{Kolekar:2012tq,
    author = "Kolekar, Sanved and Padmanabhan, T. and Sarkar, Sudipta",
    title = "{Entropy Increase during Physical Processes for Black Holes in Lanczos-Lovelock Gravity}",
    eprint = "1201.2947",
    archivePrefix = "arXiv",
    primaryClass = "gr-qc",
    doi = "10.1103/PhysRevD.86.021501",
    journal = "Phys. Rev. D",
    volume = "86",
    pages = "021501",
    year = "2012"
}

@article{Sarkar:2013swa,
    author = "Sarkar, Sudipta and Wall, Aron C.",
    title = "{Generalized second law at linear order for actions that are functions of Lovelock densities}",
    eprint = "1306.1623",
    archivePrefix = "arXiv",
    primaryClass = "gr-qc",
    doi = "10.1103/PhysRevD.88.044017",
    journal = "Phys. Rev. D",
    volume = "88",
    pages = "044017",
    year = "2013"
}

@article{Banados:1993qp,
    author = "Banados, Maximo and Teitelboim, Claudio and Zanelli, Jorge",
    title = "{Black hole entropy and the dimensional continuation of the Gauss-Bonnet theorem}",
    eprint = "gr-qc/9309026",
    archivePrefix = "arXiv",
    reportNumber = "IASSNS-HEP-93-53",
    doi = "10.1103/PhysRevLett.72.957",
    journal = "Phys. Rev. Lett.",
    volume = "72",
    pages = "957--960",
    year = "1994"
}

@article{Carlip:1994gc,
    author = "Carlip, Steven and Teitelboim, Claudio",
    title = "{Aspects of black hole quantum mechanics and thermodynamics in (2+1)-dimensions}",
    eprint = "gr-qc/9405070",
    archivePrefix = "arXiv",
    reportNumber = "IASSNS-HEP-94-34, UCD-94-13",
    doi = "10.1103/PhysRevD.51.622",
    journal = "Phys. Rev. D",
    volume = "51",
    pages = "622--631",
    year = "1995"
}

@article{Witten:2024upt,
    author = "Witten, Edward",
    title = "{Introduction to black hole thermodynamics}",
    eprint = "2412.16795",
    archivePrefix = "arXiv",
    primaryClass = "hep-th",
    doi = "10.1140/epjp/s13360-025-06288-y",
    journal = "Eur. Phys. J. Plus",
    volume = "140",
    number = "5",
    pages = "430",
    year = "2025"
}

@article{Witten:1986qs,
    author = "Witten, Edward",
    title = "{Interacting Field Theory of Open Superstrings}",
    reportNumber = "PRINT-86-0310 (PRINCETON)",
    doi = "10.1016/0550-3213(86)90298-1",
    journal = "Nucl. Phys. B",
    volume = "276",
    pages = "291--324",
    year = "1986"
}

@incollection{Crnkovic:1987tz,
    author    = "Crnkovic, C. and Witten, Edward",
    title     = "{Covariant Description of Canonical Formalism in Geometrical Theories}",
    booktitle = "{Three Hundred Years of Gravitation}",
    editor    = "Hawking, S. W. and Israel, W.",
    publisher = "Cambridge University Press",
    address   = "Cambridge",
    year      = "1987",
    pages     = "676--684"
}

@article{Ashtekar:2002qc,
    author = "Ashtekar, Abhay and Wisniewski, Jacek and Dreyer, Olaf",
    title = "{Isolated horizons in (2+1) gravity}",
    eprint = "gr-qc/0206024",
    archivePrefix = "arXiv",
    doi = "10.4310/ATMP.2002.v6.n3.a3",
    journal = "Adv. Theor. Math. Phys.",
    volume = "6",
    pages = "507--555",
    year = "2003"
}

@article{Ryu:2006bv,
    author = "Ryu, Shinsei and Takayanagi, Tadashi",
    title = "{Holographic derivation of entanglement entropy from AdS/CFT}",
    eprint = "hep-th/0603001",
    archivePrefix = "arXiv",
    reportNumber = "NSF-KITP-06-11, NSF-KITP-06-11",
    doi = "10.1103/PhysRevLett.96.181602",
    journal = "Phys. Rev. Lett.",
    volume = "96",
    pages = "181602",
    year = "2006"
}

@article{Ashtekar:1998sp,
    author = "Ashtekar, Abhay and Beetle, Christopher and Fairhurst, Stephen",
    title = "{Isolated horizons: A Generalization of black hole mechanics}",
    eprint = "gr-qc/9812065",
    archivePrefix = "arXiv",
    doi = "10.1088/0264-9381/16/2/027",
    journal = "Class. Quant. Grav.",
    volume = "16",
    pages = "L1--L7",
    year = "1999"
}

@article{Ashtekar:1999ex,
    author = "Ashtekar, Abhay",
    editor = "Kowalski-Glikman, J.",
    title = "{Classical quantum physics of isolated horizons: A Brief overview}",
    reportNumber = "CGPG-99-8-4",
    doi = "10.1007/3-540-46634-7_2",
    journal = "Lect. Notes Phys.",
    volume = "541",
    pages = "50--70",
    year = "2000"
}

@article{Ashtekar:1999yj,
    author = "Ashtekar, Abhay and Beetle, Christopher and Fairhurst, Stephen",
    title = "{Mechanics of isolated horizons}",
    eprint = "gr-qc/9907068",
    archivePrefix = "arXiv",
    doi = "10.1088/0264-9381/17/2/301",
    journal = "Class. Quant. Grav.",
    volume = "17",
    pages = "253--298",
    year = "2000"
}

@article{Ashtekar:1999wa,
    author = "Ashtekar, Abhay and Corichi, Alejandro and Krasnov, Kirill",
    title = "{Isolated horizons: The Classical phase space}",
    eprint = "gr-qc/9905089",
    archivePrefix = "arXiv",
    reportNumber = "NSF-ITP-99-33, NSF-ITP-9-33, CGPG-99-5-5",
    doi = "10.4310/ATMP.1999.v3.n3.a1",
    journal = "Adv. Theor. Math. Phys.",
    volume = "3",
    pages = "419--478",
    year = "1999"
}

@article{Ashtekar:2000sz,
    author = "Ashtekar, Abhay and Beetle, Christopher and Dreyer, Olaf and Fairhurst, Stephen and Krishnan, Badri and Lewandowski, Jerzy and Wisniewski, Jacek",
    title = "{Isolated horizons and their applications}",
    eprint = "gr-qc/0006006",
    archivePrefix = "arXiv",
    doi = "10.1103/PhysRevLett.85.3564",
    journal = "Phys. Rev. Lett.",
    volume = "85",
    pages = "3564--3567",
    year = "2000"
}

@article{Ashtekar:2004cn,
    author = "Ashtekar, Abhay and Krishnan, Badri",
    title = "{Isolated and dynamical horizons and their applications}",
    eprint = "gr-qc/0407042",
    archivePrefix = "arXiv",
    doi = "10.12942/lrr-2004-10",
    journal = "Living Rev. Rel.",
    volume = "7",
    pages = "10",
    year = "2004"
}

@article{Ashtekar:2000hw,
    author = "Ashtekar, Abhay and Fairhurst, Stephen and Krishnan, Badri",
    title = "{Isolated horizons: Hamiltonian evolution and the first law}",
    eprint = "gr-qc/0005083",
    archivePrefix = "arXiv",
    doi = "10.1103/PhysRevD.62.104025",
    journal = "Phys. Rev. D",
    volume = "62",
    pages = "104025",
    year = "2000"
}

@article{Ashtekar:1999sn,
    author = "Ashtekar, Abhay and Corichi, Alejandro",
    title = "{Laws governing isolated horizons: Inclusion of dilaton couplings}",
    eprint = "gr-qc/9910068",
    archivePrefix = "arXiv",
    reportNumber = "CGPG-99-10-2",
    doi = "10.1088/0264-9381/17/6/301",
    journal = "Class. Quant. Grav.",
    volume = "17",
    pages = "1317--1332",
    year = "2000"
}

@article{Ashtekar:2001is,
    author = "Ashtekar, Abhay and Beetle, Christopher and Lewandowski, Jerzy",
    title = "{Mechanics of rotating isolated horizons}",
    eprint = "gr-qc/0103026",
    archivePrefix = "arXiv",
    doi = "10.1103/PhysRevD.64.044016",
    journal = "Phys. Rev. D",
    volume = "64",
    pages = "044016",
    year = "2001"
}

@article{Ashtekar:2003jh,
    author = "Ashtekar, Abhay and Corichi, Alejandro and Sudarsky, Daniel",
    title = "{Nonminimally coupled scalar fields and isolated horizons}",
    eprint = "gr-qc/0305044",
    archivePrefix = "arXiv",
    reportNumber = "CGPG-2003-05-2, ICN-UNAM-03-05",
    doi = "10.1088/0264-9381/20/15/310",
    journal = "Class. Quant. Grav.",
    volume = "20",
    pages = "3413--3426",
    year = "2003"
}

@article{Robinson:1960zzb,
    author = "Robinson, Ivor and Trautman, A.",
    title = "{Spherical Gravitational Waves}",
    doi = "10.1103/PhysRevLett.4.431",
    journal = "Phys. Rev. Lett.",
    volume = "4",
    pages = "431--432",
    year = "1960"
}

@article{Banados:1992wn,
    author = "Banados, Maximo and Teitelboim, Claudio and Zanelli, Jorge",
    title = "{The Black hole in three-dimensional space-time}",
    eprint = "hep-th/9204099",
    archivePrefix = "arXiv",
    reportNumber = "PRINT-92-0151 (CHILE), IASSNS-HEP-92-29",
    doi = "10.1103/PhysRevLett.69.1849",
    journal = "Phys. Rev. Lett.",
    volume = "69",
    pages = "1849--1851",
    year = "1992"
}

@article{Ashtekar:2025wnu,
    author = "Ashtekar, Abhay and Krishnan, Badri",
    title = "{Quasi-local black hole horizons: recentadvances}",
    eprint = "2502.11825",
    archivePrefix = "arXiv",
    primaryClass = "gr-qc",
    doi = "10.1007/s41114-025-00061-4",
    journal = "Living Rev. Rel.",
    volume = "28",
    number = "1",
    pages = "8",
    year = "2025"
}

@article{Dreyer:2002mx,
    author = "Dreyer, Olaf and Krishnan, Badri and Shoemaker, Deirdre and Schnetter, Erik",
    title = "{Introduction to isolated horizons in numerical relativity}",
    eprint = "gr-qc/0206008",
    archivePrefix = "arXiv",
    doi = "10.1103/PhysRevD.67.024018",
    journal = "Phys. Rev. D",
    volume = "67",
    pages = "024018",
    year = "2003"
}

@article{Romano:1991up,
    author = "Romano, Joseph D.",
    title = "{Geometrodynamics versus connection dynamics (in the context of (2+1) and (3+1) gravity}",
    eprint = "gr-qc/9303032",
    archivePrefix = "arXiv",
    reportNumber = "PRINT-91-0319 (SYRACUSE)",
    doi = "10.1007/BF00758384",
    journal = "Gen. Rel. Grav.",
    volume = "25",
    pages = "759--854",
    year = "1993"
}

@article{Witten:1988hc,
    author = "Witten, Edward",
    title = "{(2+1)-Dimensional Gravity as an Exactly Soluble System}",
    reportNumber = "IASSNS-HEP-88-32",
    doi = "10.1016/0550-3213(88)90143-5",
    journal = "Nucl. Phys. B",
    volume = "311",
    pages = "46",
    year = "1988"
}

@article{Ashtekar:2002ag,
    author = "Ashtekar, Abhay and Krishnan, Badri",
    title = "{Dynamical horizons: Energy, angular momentum, fluxes and balance laws}",
    eprint = "gr-qc/0207080",
    archivePrefix = "arXiv",
    doi = "10.1103/PhysRevLett.89.261101",
    journal = "Phys. Rev. Lett.",
    volume = "89",
    pages = "261101",
    year = "2002"
}

@article{Zhang:1988wy,
    author = "Zhang, S. C. and Hansson, T. H. and Kivelson, S.",
    title = "{An effective field theory model for the fractional quantum hall effect}",
    reportNumber = "NSF-ITP-88-119",
    doi = "10.1103/PhysRevLett.62.82",
    journal = "Phys. Rev. Lett.",
    volume = "62",
    pages = "82--85",
    year = "1988"
}

@article{Frohlich:1990xz,
    author = "Frohlich, J. and Kerler, T.",
    title = "{Universality in quantum Hall systems}",
    reportNumber = "ETH-TH-90-26",
    doi = "10.1016/0550-3213(91)90360-A",
    journal = "Nucl. Phys. B",
    volume = "354",
    pages = "369--417",
    year = "1991"
}

@article{Balachandran:1991dw,
    author = "Balachandran, A. P. and Bimonte, G. and Gupta, K. S. and Stern, A.",
    title = "{Conformal edge currents in Chern-Simons theories}",
    eprint = "hep-th/9110072",
    archivePrefix = "arXiv",
    reportNumber = "SU-4228-487, INFN-NA-IV-91-12, UAHEP-917, UAHEP-9113",
    doi = "10.1142/S0217751X92002106",
    journal = "Int. J. Mod. Phys. A",
    volume = "7",
    pages = "4655--4670",
    year = "1992"
}

@article{Wen:1992vi,
    author = "Wen, Xiao-Gang",
    title = "{Theory of the edge states in fractional quantum Hall effects}",
    doi = "10.1142/S0217979292000840",
    journal = "Int. J. Mod. Phys. B",
    volume = "6",
    pages = "1711--1762",
    year = "1992"
}

@article{Freidel:2020xyx,
    author = "Freidel, Laurent and Geiller, Marc and Pranzetti, Daniele",
    title = "{Edge modes of gravity. Part I. Corner potentials and charges}",
    eprint = "2006.12527",
    archivePrefix = "arXiv",
    primaryClass = "hep-th",
    doi = "10.1007/JHEP11(2020)026",
    journal = "JHEP",
    volume = "11",
    pages = "026",
    year = "2020"
}

@article{Freidel:2020svx,
    author = "Freidel, Laurent and Geiller, Marc and Pranzetti, Daniele",
    title = "{Edge modes of gravity. Part II. Corner metric and Lorentz charges}",
    eprint = "2007.03563",
    archivePrefix = "arXiv",
    primaryClass = "hep-th",
    doi = "10.1007/JHEP11(2020)027",
    journal = "JHEP",
    volume = "11",
    pages = "027",
    year = "2020"
}

@article{Freidel:2020ayo,
    author = "Freidel, Laurent and Geiller, Marc and Pranzetti, Daniele",
    title = "{Edge modes of gravity. Part III. Corner simplicity constraints}",
    eprint = "2007.12635",
    archivePrefix = "arXiv",
    primaryClass = "hep-th",
    doi = "10.1007/JHEP01(2021)100",
    journal = "JHEP",
    volume = "01",
    pages = "100",
    year = "2021"
}

@article{Garg:2026mpo,
author = "Garg, Akriti and Chatterjee, Ayan",
title = "{Hawking radiation from black holes in 2+1 dimensions}",
eprint = "2604.21540",
archivePrefix = "arXiv",
primaryClass = "gr-qc",
doi = "",
journal = "",
volume = "",
number = "",
pages = "",
month = "4",
year = "2026"
}

@article{Galloway:2005mf,
    author = "Galloway, Gregory J. and Schoen, Richard",
    title = "{A Generalization of Hawking's black hole topology theorem to higher dimensions}",
    eprint = "gr-qc/0509107",
    archivePrefix = "arXiv",
    doi = "10.1007/s00220-006-0019-z",
    journal = "Commun. Math. Phys.",
    volume = "266",
    pages = "571--576",
    year = "2006"
}

@article{bib23,
  author      = "M. Ortaggio and V. Pravda and A. Pravdova",
  title       = "Ricci identities in higher dimensions",
  journal     = "Class. {Q}uant. {G}rav.",
  volume      = "24",
  pages       = "1657--1664",
  year        = "2007",
  doi         = "10.1088/0264-9381/24/6/018"
}

@article{Coley:2004jv,
    author = "Coley, A. and Milson, R. and Pravda, Vojtech and Pravdova, A.",
    title = "{Classification of the Weyl tensor in higher dimensions}",
    eprint = "gr-qc/0401008",
    archivePrefix = "arXiv",
    reportNumber = "CLASS.QUANT.GRAV.",
    doi = "10.1088/0264-9381/21/7/L01",
    journal = "Class. Quant. Grav.",
    volume = "21",
    pages = "L35--L42",
    year = "2004"
}

@article{Pravda:2004ka,
    author = "Pravda, Vojtech and Pravdova, A. and Coley, A. and Milson, R.",
    title = "{Bianchi identities in higher dimensions}",
    eprint = "gr-qc/0401013",
    archivePrefix = "arXiv",
    doi = "10.1088/0264-9381/24/6/C01",
    journal = "Class. Quant. Grav.",
    volume = "21",
    pages = "2873--2898",
    year = "2004",
    note = "[Erratum: Class.Quant.Grav. 24, 1691 (2007)]"
}

@article{Ortaggio:2009bz,
    author = "Ortaggio, Marcello and Pravda, Vojtech and Pravdova, Alena",
    editor = "Kunze, Kerstin E. and Mars, Marc and Vazquez-Mozo, Miguel Angel",
    title = "{On Kerr-Schild spacetimes in higher dimensions}",
    eprint = "0901.1561",
    archivePrefix = "arXiv",
    primaryClass = "gr-qc",
    doi = "10.1063/1.3141327",
    journal = "AIP Conf. Proc.",
    volume = "1122",
    number = "1",
    pages = "364--367",
    year = "2009"
}

@article{Ortaggio:2009zt,
    author = "Ortaggio, Marcello and Pravda, Vojtech and Pravdova, Alena",
    title = "{On asymptotically flat algebraically special spacetimes in higher dimensions}",
    eprint = "0907.1780",
    archivePrefix = "arXiv",
    primaryClass = "gr-qc",
    doi = "10.1103/PhysRevD.80.084041",
    journal = "Phys. Rev. D",
    volume = "80",
    pages = "084041",
    year = "2009"
}

@article{Ortaggio:2012cp,
    author = "Ortaggio, Marcello and Pravda, Vojtech and Pravdova, Alena",
    title = "{On the Goldberg-Sachs theorem in higher dimensions in the non-twisting case}",
    eprint = "1211.2660",
    archivePrefix = "arXiv",
    primaryClass = "gr-qc",
    doi = "10.1088/0264-9381/30/7/075016",
    journal = "Class. Quant. Grav.",
    volume = "30",
    pages = "075016",
    year = "2013"
}

@article{Ortaggio:2012jd,
    author = "Ortaggio, Marcello and Pravda, Vojtech and Pravdova, Alena",
    title = "{Algebraic classification of higher dimensional spacetimes based on null alignment}",
    eprint = "1211.7289",
    archivePrefix = "arXiv",
    primaryClass = "gr-qc",
    doi = "10.1088/0264-9381/30/1/013001",
    journal = "Class. Quant. Grav.",
    volume = "30",
    pages = "013001",
    year = "2013"
}

@article{Ortaggio:2014sea,
    author = "Ortaggio, Marcello and Pravdov{\'a}, Alena",
    title = "{Asymptotic behavior of the Weyl tensor in higher dimensions}",
    eprint = "1403.7559",
    archivePrefix = "arXiv",
    primaryClass = "gr-qc",
    doi = "10.1103/PhysRevD.90.104011",
    journal = "Phys. Rev. D",
    volume = "90",
    number = "10",
    pages = "104011",
    year = "2014"
}

@article{Devdutt:2026bmlt,
author = "Devdutt, Sahil and Garg, Akriti and Chatterjee, Ayan",
title = "{Black hole mechanics in Lovelock theory}",
eprint = "",
archivePrefix = "arXiv",
primaryClass = "gr-qc",
doi = "",
journal = "",
volume = "",
number = "",
pages = "",
year = "2026",
note = "Under review"
}

@article{Chatterjee:2015lwa,
    author = "Chatterjee, Ayan and Ghosh, Avirup",
    title = "{Quasilocal first law of black hole dynamics from local Lorentz transformations}",
    eprint = "1511.01709",
    archivePrefix = "arXiv",
    primaryClass = "gr-qc",
    doi = "10.1140/epjc/s10052-018-6021-8",
    journal = "Eur. Phys. J. C",
    volume = "78",
    number = "7",
    pages = "550",
    year = "2018"
}

@article{Zuckerman:1986vzu,
    author = "Zuckerman, Gregg J.",
    title = "{ACTION PRINCIPLES AND GLOBAL GEOMETRY}",
    reportNumber = "Print-89-0321 (YALE)",
    journal = "Conf. Proc. C",
    volume = "8607214",
    pages = "259--284",
    year = "1986"
}

@incollection{ASHTEKAR1991417,
title = {THE COVARIANT PHASE SPACE OF ASYMPTOTICALLY FLAT GRAVITATIONAL FIELDS},
editor = {Mauro Francaviglia},
booktitle = {Mechanics, Analysis and Geometry: 200 Years After Lagrange},
publisher = {Elsevier},
address = {Amsterdam},
pages = {417-450},
year = {1991},
series = {North-Holland Delta Series},
issn = {09275029},
doi = {https://doi.org/10.1016/B978-0-444-88958-4.50021-5},
url = {https://www.sciencedirect.com/science/article/pii/B9780444889584500215},
author = {Abhay Ashtekar and Luca Bombelli and Oscar Reula}
}

@book{Chandrasekhar:1985kt,
    author = "Chandrasekhar, Subrahmanyan",
    title = "{The mathematical theory of black holes}",
    isbn = "978-0-19-850370-5",
    year = "1985"
}

@book{Carmeli:1977,
author = "Carmeli, Moshe",
title = "{Group Theory and General Relativity: Representations of the Lorentz Group and Their Applications to the Gravitational Field}",
publisher = "McGraw-Hill",
address = "New York",
year = "1977"
}

@article{Milson:2004jx,
    author = "Milson, R. and Coley, A. and Pravda, Vojtech and Pravdova, A.",
    title = "{Alignment and algebraically special tensors in Lorentzian geometry}",
    eprint = "gr-qc/0401010",
    archivePrefix = "arXiv",
    doi = "10.1142/S0219887805000491",
    journal = "Int. J. Geom. Meth. Mod. Phys.",
    volume = "2",
    pages = "41--61",
    year = "2005"
}

@inproceedings{Zanelli:2005sa,
    author = "Zanelli, Jorge",
    title = "{Lecture notes on Chern-Simons (super-)gravities. Second edition (February 2008)}",
    booktitle = "{7th Mexican Workshop on Particles and Fields}",
    eprint = "hep-th/0502193",
    archivePrefix = "arXiv",
    month = "2",
    year = "2005"
}

@article{Zumino:1985dp,
    author = "Zumino, Bruno",
    title = "{Gravity Theories in More Than Four-Dimensions}",
    reportNumber = "UCB-PTH-85-13, LBL-19302",
    doi = "10.1016/0370-1573(86)90076-1",
    journal = "Phys. Rept.",
    volume = "137",
    pages = "109",
    year = "1986"
}

@article{Banados:1993ur,
    author = "Banados, Maximo and Teitelboim, Claudio and Zanelli, Jorge",
    title = "{Dimensionally continued black holes}",
    eprint = "gr-qc/9307033",
    archivePrefix = "arXiv",
    reportNumber = "IASSNS-AST-93-45",
    doi = "10.1103/PhysRevD.49.975",
    journal = "Phys. Rev. D",
    volume = "49",
    pages = "975--986",
    year = "1994"
}

@article{Corichi:2010ur,
    author = "Corichi, Alejandro and Wilson-Ewing, Edward",
    title = "{Surface terms, Asymptotics and Thermodynamics of the Holst Action}",
    eprint = "1005.3298",
    archivePrefix = "arXiv",
    primaryClass = "gr-qc",
    doi = "10.1088/0264-9381/27/20/205015",
    journal = "Class. Quant. Grav.",
    volume = "27",
    pages = "205015",
    year = "2010"
}

@article{Corichi:2016zac,
    author = "Corichi, Alejandro and Rubalcava-Garc{\'\i}a, Irais and Vuka{\v{s}}inac, Tatjana",
    title = "{Actions, topological terms and boundaries in first-order gravity: A review}",
    eprint = "1604.07764",
    archivePrefix = "arXiv",
    primaryClass = "gr-qc",
    doi = "10.1142/S0218271816300111",
    journal = "Int. J. Mod. Phys. D",
    volume = "25",
    number = "04",
    pages = "1630011",
    year = "2016"
}

@article{Corichi:2013zza,
    author = "Corichi, Alejandro and Rubalcava, Irais and Vukasinac, Tatjana",
    title = "{Hamiltonian and Noether charges in first order gravity}",
    eprint = "1312.7828",
    archivePrefix = "arXiv",
    primaryClass = "gr-qc",
    doi = "10.1007/s10714-014-1813-0",
    journal = "Gen. Rel. Grav.",
    volume = "46",
    pages = "1813",
    year = "2014"
}

@article{Corichi:2016eoe,
    author = "Corichi, Alejandro and Reyes, Juan D. and Vuka{\v{s}}inac, Tatjana",
    title = "{Weakly Isolated Horizons: First order actions and gauge symmetries}",
    eprint = "1612.01462",
    archivePrefix = "arXiv",
    primaryClass = "gr-qc",
    doi = "10.1088/1361-6382/aa631c",
    journal = "Class. Quant. Grav.",
    volume = "34",
    number = "8",
    pages = "085005",
    year = "2017"
}

@article{Basu:2010hv,
    author = "Basu, Rudranil and Chatterjee, Ayan and Ghosh, Amit",
    title = "{Local symmetries of non-expanding horizons}",
    eprint = "1004.3200",
    archivePrefix = "arXiv",
    primaryClass = "gr-qc",
    doi = "10.1088/0264-9381/29/23/235010",
    journal = "Class. Quant. Grav.",
    volume = "29",
    pages = "235010",
    year = "2012"
}

@article{Wall:2010cj,
    author = "Wall, Aron C.",
    title = "{A Proof of the generalized second law for rapidly-evolving Rindler horizons}",
    eprint = "1007.1493",
    archivePrefix = "arXiv",
    primaryClass = "gr-qc",
    doi = "10.1103/PhysRevD.82.124019",
    journal = "Phys. Rev. D",
    volume = "82",
    pages = "124019",
    year = "2010"
}

@article{Chatterjee:2008if,
    author = "Chatterjee, Ayan and Ghosh, Amit",
    title = "{Laws of Black Hole Mechanics from Holst Action}",
    eprint = "0812.2121",
    archivePrefix = "arXiv",
    primaryClass = "gr-qc",
    doi = "10.1103/PhysRevD.80.064036",
    journal = "Phys. Rev. D",
    volume = "80",
    pages = "064036",
    year = "2009"
}

@article{Jacobson:2015uqa,
    author = "Jacobson, Ted and Mohd, Arif",
    title = "{Black hole entropy and Lorentz-diffeomorphism Noether charge}",
    eprint = "1507.01054",
    archivePrefix = "arXiv",
    primaryClass = "gr-qc",
    doi = "10.1103/PhysRevD.92.124010",
    journal = "Phys. Rev. D",
    volume = "92",
    pages = "124010",
    year = "2015"
}

@article{Horndeski:1974wa,
    author = "Horndeski, Gregory Walter",
    title = "{Second-order scalar-tensor field equations in a four-dimensional space}",
    doi = "10.1007/BF01807638",
    journal = "Int. J. Theor. Phys.",
    volume = "10",
    pages = "363--384",
    year = "1974"
}

@article{Lovelock:1971yv,
    author = "Lovelock, D.",
    title = "{The Einstein tensor and its generalizations}",
    doi = "10.1063/1.1665613",
    journal = "J. Math. Phys.",
    volume = "12",
    pages = "498--501",
    year = "1971"
}

@article{Troncoso:1999pk,
    author = "Troncoso, Ricardo and Zanelli, Jorge",
    title = "{Higher dimensional gravity, propagating torsion and AdS gauge invariance}",
    eprint = "hep-th/9907109",
    archivePrefix = "arXiv",
    reportNumber = "CECS-PH-99-06",
    doi = "10.1088/0264-9381/17/21/307",
    journal = "Class. Quant. Grav.",
    volume = "17",
    pages = "4451--4466",
    year = "2000"
}

@article{Chatterjee:2025jur,
    author = "Chatterjee, Ayan and Devdutt, Sahil and Ghosh, Avirup",
    title = "{Laws of black hole mechanics in the Einstein-Gauss-Bonnet theory}",
    eprint = "2510.24189",
    archivePrefix = "arXiv",
    primaryClass = "gr-qc",
    doi = "10.1007/s10714-026-03526-1",
    journal = "Gen. Rel. Grav.",
    volume = "58",
    number = "3",
    pages = "25",
    year = "2026"
}

@article{Krishnan:2009tj,
    author = "Krishnan, Chethan and Kuperstein, Stanislav",
    title = "{A Comment on Kerr-CFT and Wald Entropy}",
    eprint = "0903.2169",
    archivePrefix = "arXiv",
    primaryClass = "hep-th",
    doi = "10.1016/j.physletb.2009.05.056",
    journal = "Phys. Lett. B",
    volume = "677",
    pages = "326--331",
    year = "2009"
}

@article{Dong:2013qoa,
    author = "Dong, Xi",
    title = "{Holographic Entanglement Entropy for General Higher Derivative Gravity}",
    eprint = "1310.5713",
    archivePrefix = "arXiv",
    primaryClass = "hep-th",
    reportNumber = "SU-ITP-13-21",
    doi = "10.1007/JHEP01(2014)044",
    journal = "JHEP",
    volume = "01",
    pages = "044",
    year = "2014"
}

@article{Bhattacharyya:2021jhr,
    author = "Bhattacharyya, Sayantani and Dhivakar, Prateksh and Dinda, Anirban and Kundu, Nilay and Patra, Milan and Roy, Shuvayu",
    title = "{An entropy current and the second law in higher derivative theories of gravity}",
    eprint = "2105.06455",
    archivePrefix = "arXiv",
    primaryClass = "hep-th",
    doi = "10.1007/JHEP09(2021)169",
    journal = "JHEP",
    volume = "09",
    pages = "169",
    year = "2021"
}

@article{Hollands:2024vbe,
    author = "Hollands, Stefan and Wald, Robert M. and Zhang, Victor G.",
    title = "{Entropy of dynamical black holes}",
    eprint = "2402.00818",
    archivePrefix = "arXiv",
    primaryClass = "hep-th",
    doi = "10.1103/PhysRevD.110.024070",
    journal = "Phys. Rev. D",
    volume = "110",
    number = "2",
    pages = "024070",
    year = "2024"
}

@article{Bhattacharyya:2026aiu,
    author = "Bhattacharyya, Sayantani and Biswas, Parthajit and Kundu, Nilay",
    title = "{A comparison of two constructions for dynamical corrections to Wald entropy}",
    eprint = "2607.27732",
    archivePrefix = "arXiv",
    primaryClass = "hep-th",
    doi = "",
    journal = "arXiv",
    volume = "",
    number = "",
    pages = "",
    month = "7",
    year = "2026"
}

@article{Jacobson:1993vj,
    author = "Jacobson, Ted and Kang, Gungwon and Myers, Robert C.",
    title = "{On black hole entropy}",
    eprint = "gr-qc/9312023",
    archivePrefix = "arXiv",
    reportNumber = "MCGILL-93-22, NSF-ITP-93-152, UMDGR-94-75",
    doi = "10.1103/PhysRevD.49.6587",
    journal = "Phys. Rev. D",
    volume = "49",
    pages = "6587--6598",
    year = "1994"
}

@article{Sen:2005wa,
    author = "Sen, Ashoke",
    title = "{Black hole entropy function and the attractor mechanism in higher derivative gravity}",
    eprint = "hep-th/0506177",
    archivePrefix = "arXiv",
    doi = "10.1088/1126-6708/2005/09/038",
    journal = "JHEP",
    volume = "09",
    pages = "038",
    year = "2005"
}

@article{Astefanesei:2006dd,
    author = "Astefanesei, Dumitru and Goldstein, Kevin and Jena, Rudra P. and Sen, Ashoke and Trivedi, Sandip P.",
    title = "{Rotating attractors}",
    eprint = "hep-th/0606244",
    archivePrefix = "arXiv",
    reportNumber = "TIFR-TH-06-15, HRI-P-06-06-002",
    doi = "10.1088/1126-6708/2006/10/058",
    journal = "JHEP",
    volume = "10",
    pages = "058",
    year = "2006"
}

@article{Hubeny:2007xt,
    author = "Hubeny, Veronika E. and Rangamani, Mukund and Takayanagi, Tadashi",
    title = "{A Covariant holographic entanglement entropy proposal}",
    eprint = "0705.0016",
    archivePrefix = "arXiv",
    primaryClass = "hep-th",
    reportNumber = "DCPT-07-13, KUNS-2069",
    doi = "10.1088/1126-6708/2007/07/062",
    journal = "JHEP",
    volume = "07",
    pages = "062",
    year = "2007"
}

@article{Camps:2013zua,
    author = "Camps, Joan",
    title = "{Generalized entropy and higher derivative Gravity}",
    eprint = "1310.6659",
    archivePrefix = "arXiv",
    primaryClass = "hep-th",
    doi = "10.1007/JHEP03(2014)070",
    journal = "JHEP",
    volume = "03",
    pages = "070",
    year = "2014"
}

@article{Faulkner:2013ana,
    author = "Faulkner, Thomas and Lewkowycz, Aitor and Maldacena, Juan",
    title = "{Quantum corrections to holographic entanglement entropy}",
    eprint = "1307.2892",
    archivePrefix = "arXiv",
    primaryClass = "hep-th",
    doi = "10.1007/JHEP11(2013)074",
    journal = "JHEP",
    volume = "11",
    pages = "074",
    year = "2013"
}

@article{Nishioka:2009un,
    author = "Nishioka, Tatsuma and Ryu, Shinsei and Takayanagi, Tadashi",
    title = "{Holographic Entanglement Entropy: An Overview}",
    eprint = "0905.0932",
    archivePrefix = "arXiv",
    primaryClass = "hep-th",
    reportNumber = "KUNS-2207, IPMU09-0056, KUNS-2207, IPMU09-0056",
    doi = "10.1088/1751-8113/42/50/504008",
    journal = "J. Phys. A",
    volume = "42",
    pages = "504008",
    year = "2009"
}

@article{VanRaamsdonk:2016exw,
    author = "Van Raamsdonk, Mark",
    title = "{Lectures on gravity and entanglement.}",
    eprint = "1609.00026",
    archivePrefix = "arXiv",
    primaryClass = "hep-th",
    doi = "10.1142/9789813149441_0005",
    journal = " ",
    pages = "297--351",
    year = "2017"
}

\end{document}